\documentclass{article}

\usepackage[preprint]{neurips_2026}

\usepackage[utf8]{inputenc}
\usepackage[T1]{fontenc}
\usepackage{hyperref}
\usepackage{url}
\usepackage{booktabs}
\usepackage{amsmath,amssymb,amsfonts}
\usepackage{nicefrac}
\usepackage{microtype}
\usepackage{xcolor}
\usepackage{graphicx}
\usepackage{tabularx}
\usepackage{mdframed}
\usepackage{placeins}   
\usepackage[capitalise,nameinlink]{cleveref}

\setcitestyle{authoryear,square,comma}
\hypersetup{colorlinks=true, linkcolor=blue!60!black, citecolor=blue!60!black, urlcolor=blue!60!black}

\newmdenv[backgroundcolor=black!4, linecolor=black!25, linewidth=0.4pt, innerleftmargin=8pt,
  innerrightmargin=8pt, innertopmargin=6pt, innerbottommargin=6pt, skipabove=6pt, skipbelow=6pt]{takeaway}

\newcommand{\file}[1]{\texttt{#1}}

\newcommand{\snapshotDate}{2026-10-16}  
\newcommand{\minStars}{1,000}  
\newcommand{\activeDays}{90}  
\newcommand{\activeSince}{2026-07-18}  
\newcommand{\nPopulation}{22,854}  
\newcommand{\nAdopters}{6,006}  
\newcommand{\pctAdopters}{26.3\%}  
\newcommand{\thinImportChars}{300}  
\newcommand{\commitSample}{100}  
\newcommand{\metricsDays}{365}  
\newcommand{\thresholdLow}{100}  
\newcommand{\thresholdHigh}{2,000}  
\newcommand{\nHistoryFiles}{8,393}  
\newcommand{\nDocuments}{5,998}  
\newcommand{\nFilesB}{6,578}  
\newcommand{\nTypesB}{16}  
\newcommand{\nGold}{324}  
\newcommand{\nGoldHalf}{162}  
\newcommand{\kappaModelsB}{0.67--0.80}  
\newcommand{\kappaHumanB}{0.46--0.52}  
\newcommand{\costB}{\$13.83}  
\newcommand{\nDiscoveryDocs}{350}  
\newcommand{\nDiscoveryAgents}{14}  
\newcommand{\nSaturationDocs}{80}  
\newcommand{\nDimensions}{10}  
\newcommand{\nCodebookValues}{48}  
\newcommand{\nFunctions}{14}  
\newcommand{\nStatements}{341,852}  
\newcommand{\nFilesSegmented}{6,577}  
\newcommand{\pctVerbatim}{99.4\%}  
\newcommand{\nSubjects}{57}  
\newcommand{\nAreas}{10}  
\newcommand{\nTaxonomySample}{2,500}  
\newcommand{\nTaxonomyCheck}{1,500}  
\newcommand{\pctTaxonomyOther}{0.6\%}  
\newcommand{\nSampleFiles}{1,200}  
\newcommand{\nSampleFilesCoded}{1,198}  
\newcommand{\nSampleStatements}{61,772}  
\newcommand{\nValFiles}{43}  
\newcommand{\nValStatements}{2,151}  
\newcommand{\kappaStrongModels}{0.76--0.89}  
\newcommand{\kappaHaikuSonnet}{0.47--0.81}  
\newcommand{\pctCutOk}{92\%}  
\newcommand{\pctParaphraseOk}{93\%}  
\newcommand{\pctSubjectFits}{95\%}  
\newcommand{\kappaHumanKind}{0.33--0.40}  
\newcommand{\pctHumanDescRule}{82--83\%}  
\newcommand{\pctHumanArea}{43--50\%}  
\newcommand{\pctHumanSubject}{32--37\%}  
\newcommand{\nRevisedAfterReveal}{7}  
\newcommand{\nHaikuCoded}{339,810}  
\newcommand{\nGapFiles}{75}  
\newcommand{\nBoot}{1,000}  
\newcommand{\nHumanDrawn}{100}  
\newcommand{\nHumanCoded}{60}  
\newcommand{\elementPrecision}{0.8}  
\newcommand{\nElementSample}{1,500}  
\newcommand{\costC}{\$195.68}  
\newcommand{\nAgentsOnly}{2,566}  
\newcommand{\nClaudeOnly}{1,052}  
\newcommand{\nBoth}{2,388}  
\newcommand{\pctTierLow}{23.2\%}  
\newcommand{\pctTierTop}{59.6\%}  
\newcommand{\nTierTop}{89}  
\newcommand{\pctTypeScript}{40.8\%}  
\newcommand{\pctLangC}{13.0\%}  
\newcommand{\pctAnyFormat}{29.1\%}  
\newcommand{\nOnlyOther}{646}  
\newcommand{\nDefers}{1,540}  
\newcommand{\nReverse}{182}  
\newcommand{\nSymCA}{629}  
\newcommand{\nSymAC}{155}  
\newcommand{\nImports}{872}  
\newcommand{\nTwoReal}{598}  
\newcommand{\nTwoRealAtHighThreshold}{450}  
\newcommand{\nTwoRealAtLowThreshold}{775}  
\newcommand{\pctRenamedFromClaude}{3.8\%}  
\newcommand{\ciRenamed}{2.6--5.3}  
\newcommand{\nRenameSample}{800}  
\newcommand{\pctSameCommit}{41.8\%}  
\newcommand{\pctClaudeFirst}{35.1\%}  
\newcommand{\pctAgentsFirst}{22.8\%}  
\newcommand{\fallbackDate}{2026-09-18}  
\newcommand{\fallbackWindowDays}{29}  
\newcommand{\pctFallbackBefore}{60.8\%}  
\newcommand{\pctFallbackAfter}{68.6\%}  
\newcommand{\pctAgentsMulti}{80.0\%}  
\newcommand{\pctClaudeMulti}{67.7\%}  
\newcommand{\pctAICreatedAgents}{13.5\%}  
\newcommand{\pctAICreatedClaude}{22.7\%}  
\newcommand{\pctAICommits}{24.5\%}  
\newcommand{\nAgentFileCommits}{106,279}  
\newcommand{\medianWords}{818}  
\newcommand{\pTenWords}{163}  
\newcommand{\pNinetyWords}{3,010}  
\newcommand{\pctClaudeInit}{11.5\%}  
\newcommand{\nPRFiles}{1,345}  
\newcommand{\nCommitFiles}{1,739}  
\newcommand{\nTestFiles}{2,586}  
\newcommand{\nReviewFiles}{401}  
\newcommand{\commitConv}{36.7}  
\newcommand{\testTrivialPro}{8.1}  
\newcommand{\testWeakenPro}{4.9}  
\newcommand{\reviewStyleReq}{7.0}  
\newcommand{\reviewStylePro}{7.7}  
\newcommand{\prTestPlan}{27.2}  
\newcommand{\nClaudeInit}{692}  
\newcommand{\nCodexInit}{146}  
\newcommand{\nNextjsBlock}{44}  
\newcommand{\nCopiedRuleSet}{32}  
\newcommand{\nOriginHist}{5,942}  
\newcommand{\pctTextTrace}{19.4\%}  
\newcommand{\pctAnyTrace}{33.1\%}  
\newcommand{\pctInitSig}{15.8\%}  
\newcommand{\nInitSig}{948}  
\newcommand{\nToolBlock}{93}  
\newcommand{\pctToolInserted}{1.6\%}  
\newcommand{\pctSharedText}{3.4\%}  
\newcommand{\nSharedOnly}{104}  
\newcommand{\pctSharedOnly}{1.7\%}  
\newcommand{\pctCrossDup}{1.0\%}  
\newcommand{\pctOriginAICreated}{17.7\%}  
\newcommand{\pctOriginAIEdit}{36.4\%}  
\newcommand{\pctOriginAnyAI}{41.7\%}  
\newcommand{\pctOriginChanged}{81.8\%}  
\newcommand{\pctGenChanged}{82.4\%}  
\newcommand{\pctNoGenChanged}{81.7\%}  
\newcommand{\pctGenLow}{17.1\%}  
\newcommand{\pctGenTop}{9.4\%}  
\newcommand{\nMigrations}{396}  
\newcommand{\nMigOwn}{337}  
\newcommand{\pctMigSameCommit}{66.9\%}  
\newcommand{\pctMigAI}{19.9\%}  
\newcommand{\pctMigAIBase}{26.2\%}  
\newcommand{\pctClaudeOfAIUsers}{88.0\%}  
\newcommand{\paraMinChars}{80}  
\newcommand{\sharedOwners}{3}  
\newcommand{\nearDupJaccard}{0.8}  
\newcommand{\hOneAgentIRR}{0.977}  
\newcommand{\hFourOR}{0.93}  
\newcommand{\hThreeOR}{1.68}  
\newcommand{\hSixOR}{1.36}  
\newcommand{\nAIBanRepos}{141}  
\newcommand{\pctClaudeFirstEarly}{73.7\%}  
\newcommand{\pctClaudeFirstLate}{17.1\%}  
\newcommand{\pctClaudeFirstLateMax}{21\%}  
\newcommand{\nReposEone}{992}  
\newcommand{\pctReposRule}{98.2}  
\newcommand{\ciReposRule}{97.3--98.9}  
\newcommand{\meanRuleShare}{52.6}  
\newcommand{\meanDescShare}{29.8}  
\newcommand{\meanRefShare}{17.6}  
\newcommand{\repoRulesCode}{88.7}  
\newcommand{\repoRulesWorkflow}{85.4}  
\newcommand{\repoRulesAgent}{79.3}  
\newcommand{\repoRulesFiles}{52.7}  
\newcommand{\areaCode}{74.4}  
\newcommand{\areaBuild}{72.6}  
\newcommand{\areaWorkflow}{67.3}  
\newcommand{\areaKnowledgeRule}{54.3}  
\newcommand{\areaKnowledgeAny}{86.2}  
\newcommand{\subjVerif}{49.7}  
\newcommand{\subjTestWriting}{48.2}  
\newcommand{\subjAgentFiles}{44.3}  
\newcommand{\pctReposMandatory}{80.9\%}  
\newcommand{\pctReposProhibit}{77.8\%}  
\newcommand{\meanProhibShare}{15.8\%}  
\newcommand{\pctReposLifecycle}{54.7\%}  
\newcommand{\pctReposException}{53.7\%}  
\newcommand{\pctReposReason}{82.9\%}  
\newcommand{\pctDirectivesNoReason}{74.0\%}  
\newcommand{\pctReposIncident}{10.8\%}  
\newcommand{\pctNameAgent}{8.6\%}  
\newcommand{\pctAddressHumans}{8.3\%}  
\newcommand{\pctReposVague}{40.5\%}  
\newcommand{\pctVagueShare}{3.5\%}  
\newcommand{\pctRepoAIPro}{7.5\%}  
\newcommand{\ciRepoAIPro}{6.3--8.9}  
\newcommand{\nRepoAIReq}{59}  
\newcommand{\pctRepoPRAIReq}{5.8\%}  
\newcommand{\pctRepoPRAIPro}{1.9\%}  
\newcommand{\kappaTypeSonnet}{0.82}  
\newcommand{\kappaTypePI}{0.43}  
\newcommand{\kappaAIPI}{0.80}  
\newcommand{\nSpotScored}{29}  
\newcommand{\pctAIAdopters}{27.9\%}  
\newcommand{\pctAIControls}{13.3\%}  
\newcommand{\orAIFocus}{2.76}  
\newcommand{\orLibrary}{0.58}  
\newcommand{\orDocs}{0.50}  
\newcommand{\nTestsE}{99}  
\newcommand{\nSigE}{49}  
\newcommand{\orVcCommits}{1.50}  
\newcommand{\orKnowledgeCommits}{0.77}  
\newcommand{\orTestingAge}{1.37}  
\newcommand{\nSonnetE}{927}  
\newcommand{\nReposE}{5,586}  
\newcommand{\nAdoptersCompare}{5,286}  
\newcommand{\orAIBoth}{1.54}  
\newcommand{\ciAIBoth}{1.28--1.84}  
\newcommand{\ciKappaTypePI}{0.23--0.63}  
\newcommand{\nControlsCompare}{4,699}  
\newcommand{\priorMohsenAdoption}{5\%}  
\newcommand{\priorDiazOverall}{21.7\%}  
\newcommand{\minStatementsD}{5}  
\newcommand{\nReposD}{5,587}  
\newcommand{\nTestsD}{11}  
\newcommand{\nPairsD}{484}  

\title{What Open-Source Projects Tell Their Coding Agents:\\ A Census of \texttt{AGENTS.md} and \texttt{CLAUDE.md}}
\author{%
  Pawel Zarzycki
}

\begin{document}
\maketitle
\begin{abstract}
Coding agents can be guided with instruction files in a repository. We examine this emerging practice through a census
of \nPopulation{} popular, active GitHub repositories in thirteen languages, treating each repository's root
\file{AGENTS.md} and \file{CLAUDE.md} as one instruction set. Agent instructions appear in \pctAdopters{} of
repositories and \pctTierTop{} of the most-starred; where both files exist, \file{CLAUDE.md} usually points to
\file{AGENTS.md}. We analyse \nStatements{} statements using model-assisted annotation and human review.
The annotations reveal a practical mix of project knowledge and rules about code, building and workflow.
Collaboration rules name concrete requirements, from test plans to commit formats, yet
\pctDirectivesNoReason{} of directives give no reason. Projects disagree on AI attribution: commit-message
rules more often prohibit it than require it, while pull-request rules more often require disclosure.
A third of instruction sets have a generator-associated signature, a tool-inserted block, shared text or an
AI-marked creation commit; four in five are edited after creation. Prespecified tests associate repository
popularity with rules on pull requests and AI contributions. Exploratory analyses link activity to process-oriented content and show that LLM and agent projects
are over-represented among adopters. The models agree much more closely with each other than with the human
reviewer's topic assignments, a caution for automated annotation. We release the annotations, taxonomies and
pointers to the analysed file versions.

\end{abstract}
\section{Introduction}\label{sec:introduction}

Human contributors learn a project's norms through review: which commands to run, what a pull request must
contain, when a change is done. Coding agents, which now work in open-source repositories at scale
\citep{robbes2026agenticmuch,li2026aidev}, often start a task without that history, so projects write the norms down for them, mostly in
two root files: \file{AGENTS.md}, an open format that calls itself a README for agents \citep{agentsmd2026}, and
\file{CLAUDE.md}, which Claude Code reads \citep{anthropic2026memory}. Whether such files help agents is contested
\citep{gloaguen2026evaluating,lulla2026efficiency}, so it matters what projects put in them. Existing studies
characterise them on repositories already tied to agent use \citep{chatlatanagulchai2026agentreadmes} or screen
broad samples under other definitions, with adoption estimates from \priorMohsenAdoption{} of mature repositories
\citep{mohsenimofidi2026contexteng} to \priorDiazOverall{} of agent-using ones \citep{diaz2026methharness}. What
popular projects write, statement by statement, where that text comes from and what it varies with is not known.

We answer these questions with a census of \nPopulation{} public GitHub repositories returned by GitHub's search API in thirteen languages with
at least \minStars{} stars and recent activity, and treat each repository's \file{AGENTS.md} and \file{CLAUDE.md} as
one \emph{instruction set}. We ask who has agent instructions (RQ1), what they contain (RQ2), how instructions are
expressed (RQ3), where their content comes from (RQ4) and what it varies with (RQ5).
Analysing the files together, down to individual statements, lets us examine both the requirements projects give
their agents and the traces of generators, tools and shared text in those requirements. We combine this account
with adoption and file-history measurements, prespecified hypothesis tests and exploratory associations with
project characteristics. We release file-version pointers, annotations, taxonomies and aggregate results,
without redistributing the corpus texts.

\section{Related work}\label{sec:related}

\paragraph{Agent instruction files.} \citet{chatlatanagulchai2026agentreadmes} coded the files of AIDev repositories
into a taxonomy of instruction types, and \citet{chatlatanagulchai2025manifests} and \citet{santos2026decoding}
studied \file{CLAUDE.md} files on their own. \citet{mohsenimofidi2026contexteng} found context files in few mature
repositories and identified several instruction styles; \citet{galster2026configuring,galster2026harness}
catalogued the configuration mechanisms of current agents and observed a bottom-up convergence on \file{AGENTS.md};
\citet{cardoen2026ghagentfiles} and \citet{chakrabarti2026growing} followed files over time; \citet{santos2026smells}
defined configuration smells; and Cursor rules have been studied separately
\citep{jiang2026cursorrules,sun2026cursorrulesfiles}. We differ in population and unit: a census of popular, active
repositories not selected for agent use, with each repository's files analysed together at the level of statements.

\paragraph{Do instructions help?} Controlled studies, mostly preprints, report mixed effects of instruction files on
agents' success and cost \citep{gloaguen2026evaluating,lulla2026efficiency,khatri2026ablation}, and observational
work links file features to merge rates \citep{arabat2026instructions}. Adherence varies little with layout
\citep{mcmillan2026adherence}. Preprints report that negative constraints help, although random rules helped as much
as curated ones \citep{zhang2026guardrails}, and that rules naming concrete commands or paths carry most of the
gains in agent skills \citep{wang2026skillevolution}. We make no effect claims but measure, across a population,
features this work names: prohibitions, concrete requirements, rationale and conditions.

\paragraph{Rule taxonomies.} Taxonomies of agent rules classify topics and levels \citep{cai2026ruletaxonomy},
polarity \citep{zhang2026guardrails}, security rules \citep{yan2026donot} and values guidance \citep{treude2026ethics}.
Our codebook adds how statements instruct (binding force, conditions, rationale, addressee, actionability) and, for
collaboration subjects, which concrete elements rules require.

\paragraph{AI contributions and attribution.} Mining studies measure agent activity through commit trailers and pull
requests and warn that the trace is incomplete and easily misread
\citep{robbes2026agenticmuch,robbes2026perils,khosravani2026census,yu2026simpson}. Projects increasingly publish
policies on AI contributions \citep{gao2026autopilot,hora2026aipolicy,chen2026visible}, and developers sometimes admit
AI use in commits \citep{xiao2026selfadmitted,tufano2026selfadmitted}. Agent instructions are where such policy is
written for the agent itself.

\paragraph{Repository meta-files and LLM coding.} Conventionally named files such as READMEs, contribution guidelines
and codes of conduct are an established object of mining research
\citep{prana2019readme,elazhary2019guidelines,tourani2017coc}, including how templates spread and diverge
\citep{hata2021metamaintenance}; prompts in code have been studied as artefacts too \citep{pister2024promptset}. We
follow this tradition in tracing where instruction text comes from. Language models used as annotators can beat crowd
workers in accuracy and are more consistent than human coders \citep{gilardi2023chatgpt}, but on classification they
often reach only fair agreement with humans \citep{ziems2024css}.

\section{Data and methods}\label{sec:methods}


\subsection{Population, census and collection}\label{sec:population}

We enumerated public GitHub repositories that, on \snapshotDate, had at least \minStars{} stars, had
been pushed to within the preceding \activeDays{} days, were neither forks nor archived, and had one of thirteen
primary languages: Python, JavaScript, TypeScript, Java, C\#, C++, C, Go, Rust, PHP, Ruby, Kotlin or Swift.
GitHub's search API returned \nPopulation{} such repositories. We checked every one of them for a root
\file{AGENTS.md} or \file{CLAUDE.md}, exact case, on the default branch, and call a repository with one an
\emph{adopter}. The frame is defined by popularity and activity, rather than prior evidence of agent use.

We collected files, metadata, activity and file histories through GitHub's GraphQL and REST APIs without cloning.
Adopter files were read at the snapshot commit. A commit is \emph{AI-marked} if its message or author e-mail
carries a coding agent's co-author trailer or signature. Such markers capture only part of AI-assisted work,
because tools and projects differ in whether they add them.

The object of study is a repository's \emph{instruction set}: everything its root \file{AGENTS.md} and
\file{CLAUDE.md} say, taken together, with symbolic links and thin imports resolved so that no text counts twice.
Structural features use the joined text of each set (\nDocuments{} \emph{documents}); content is coded per file
(\nFilesB{} files that do not merely defer to the other) and pooled per repository.\label{sec:units}

\subsection{Document structure and an existing taxonomy}\label{sec:stage-a}\label{sec:stage-b}

We measured document size, Markdown structure, directive words, named tools and agents, and generator-associated
signatures. For comparison with earlier work, we applied the
\nTypesB{} instruction types of \citet{chatlatanagulchai2026agentreadmes}, calibrated against that study's \nGold{}
human-labelled files and corrected for over-labelling \citep{rogan1978prevalence}, so that our population can be
compared with theirs (\cref{sec:app-stage-b}).

\subsection{Statement extraction and annotation}\label{sec:stage-c}

\paragraph{Categories.} We developed a codebook through model-assisted open coding of \nDiscoveryDocs{}
documents sampled by language, star tier and file composition. We consolidated the proposed codes and checked
for missing categories in two rounds on \nSaturationDocs{} unseen documents, followed by critique and revision.
The resulting codebook has \nDimensions{} dimensions and
\nCodebookValues{} values. A statement's \emph{function} distinguishes descriptions, references and rules, and
whether a rule governs code, workflow, the agent or the instruction files.

\paragraph{Statements.} We used Haiku to segment files into \emph{statements}, each expressing one claim or
instruction regardless of line boundaries. Each statement carries its
verbatim spans, a self-contained restatement, a free-text subject and its \emph{elements}: the parts it lists or
requires. This gave \nStatements{} statements, \pctVerbatim{} with spans found verbatim. A \emph{directive} is a
statement with binding force, from a mandatory rule to a permission; descriptions and plain references are not
directives. We also developed a subject taxonomy and, for four collaboration subjects, element categories and
their \emph{stance}: required, prohibited or mentioned. The model-assisted procedure is in \cref{sec:app-m-stage-c}.

\paragraph{Coding and estimation.} We coded statements in context for subject and all codebook dimensions, using
Haiku for the corpus and Sonnet for a stratified sample (\nSampleFilesCoded{} files coded). The sample carries the headline
estimates, because Sonnet and Opus agree closely ($\kappa$ \kappaStrongModels{} by dimension) while Haiku agrees less
with Sonnet (\kappaHaikuSonnet; \citealp{cohen1960kappa}). For a code $c$ the share of repositories whose
instruction set contains it is
\begin{equation}\label{eq:share}
  \hat{p}_c \;=\; \frac{\sum_{r \in S} w_r\, \mathbf{1}[c \in r]}{\sum_{r \in S} w_r}, \qquad w_r = \frac{1}{\pi_r},
\end{equation}
over the \nReposEone{} sample repositories $S$ whose files Sonnet coded completely, with $\pi_r$ the repository's
inclusion probability (the sample was drawn by file), and intervals from a weighted cluster bootstrap over
repositories. Mean shares per repository are computed in the same way. The element tables and the correlations use
Haiku's codes for all files, uncorrected; calibrated census estimates are in \cref{sec:app-methods}.

\paragraph{Human review.} A human reviewer (the author) coded \nHumanCoded{} statements before seeing model
annotations, then checked their boundaries, restatements, extracted elements and subjects. These came from a
stratified draw from validation files with fully matching source spans. The reviewer accepted most sampled
boundaries, restatements and extracted elements, but agreement on subject categories was limited. These checks
do not validate every subsequent category or stance assignment; the procedure and detailed results are in
Appendix~\ref{sec:review-results}.

\subsection{Prespecified tests and exploratory associations}\label{sec:stage-d}\label{sec:stage-e}

We related repository-level content, from raw Haiku codes, to project characteristics.
Counts are modelled with negative binomial regression,
\begin{equation}\label{eq:model}
  \log \mathbb{E}[Y_r] \;=\; \log m_r + \boldsymbol{\beta}^{\top}\mathbf{x}_r + \boldsymbol{\gamma}^{\top}\mathbf{z}_r ,
\end{equation}
where $Y_r$ counts statements of one kind in repository $r$ (rules, prohibitions, \ldots), $m_r$ is its reference
total (statements or rules), $\mathbf{x}_r$ are the standardised predictors under test and $\mathbf{z}_r$ the
controls: age, language and star tier, except when the latter two encode the predictor under test. Exploratory
models instead include stars as a predictor and add project type. Continuous predictors are standardised after
transformation. Indicators, such as whether an instruction set covers an area, use the logit with the log number of
statements among the controls. Standard errors are
heteroskedasticity-robust (HC1), and the false discovery rate is controlled by the Benjamini--Hochberg procedure
\citep{benjamini1995fdr} within each family.

We committed six hypotheses and their analysis plan before fitting the hypothesis-testing models, after
inspecting descriptive results. The paired file comparison was informed by a pooled difference already seen.
The tests use \nReposD{} repositories with at
least \minStatementsD{} coded statements (\nTestsD{} tests). A hypothesis counts as supported only if all its tests
have the predicted direction and an adjusted $p < 0.05$. Exploratory analyses relate the content of \nReposE{} instruction sets to
project type, language, stars, human authors, activity, age and the AI-marked share (\nTestsE{} tests), and
compare adopters with \nControlsCompare{} repositories without agent files, matched by star tier and language.
We classified project type using Haiku on repository metadata alone. Specifications and sensitivity procedures
are in \cref{sec:app-m-stage-d,sec:app-m-stage-e}; deviations are in \cref{sec:app-deviations}.

\subsection{Language models as instruments}\label{sec:llm-use}

We used Sonnet for initial code discovery and Opus for consolidation in Claude Code. Segmentation, coding and
classification used Anthropic's Claude Haiku 5.5,
Sonnet 5.5 and Opus 5.5 through Amazon Bedrock in October 2026, with structured output constrained to the codes and
the instruction to treat every file as data. Coding cost \costB{} for the existing-taxonomy comparison and
\costC{} for statement extraction and annotation. Further settings are in \cref{sec:app-m-llm-use}.

\section{Results}\label{sec:results}


\subsection{RQ1: Who has agent instructions?}\label{sec:rq1}

\begin{takeaway}
One in four popular, active repositories gives its coding agents written instructions, and adoption rises with
popularity in every tier. Where both file names exist, \file{CLAUDE.md}
usually points to \file{AGENTS.md}.
\end{takeaway}

\begin{figure}[t]
\centering
\includegraphics[width=\linewidth]{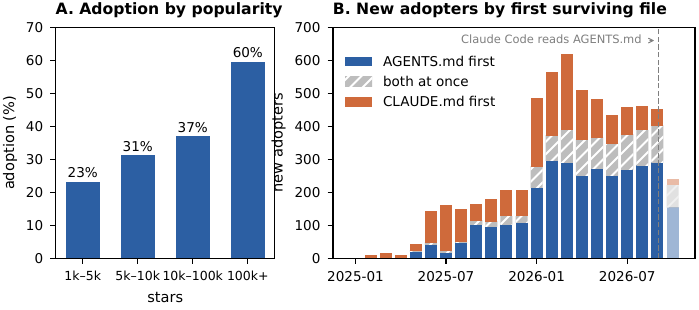}
\caption{Adoption rises with popularity, and new adopters increasingly start with \file{AGENTS.md}. \textbf{(A)} Share of repositories with agent instructions by star tier (census of \nPopulation{}).
\textbf{(B)} New adopters per month by their first surviving agent file; the dashed line marks Claude Code's support
for \file{AGENTS.md} (\fallbackDate); the last bar is a partial month.}
\label{fig:adoption}
\end{figure}

Of the \nPopulation{} repositories, \nAdopters{} (\pctAdopters) have a root \file{AGENTS.md}, a root
\file{CLAUDE.md}, or both. Adoption rises from \pctTierLow{} in the lowest star tier to \pctTierTop{} of the
\nTierTop{} repositories with the most stars (\cref{fig:adoption}A), and ranges by language from \pctTypeScript{}
of TypeScript to \pctLangC{} of C repositories (\cref{tab:languages}). Counting the other instruction formats we
listed, \pctAnyFormat{} have at least one; only \nOnlyOther{} repositories use another format without a root file.

\paragraph{One instruction set, two file names.} \nAgentsOnly{} adopters have only \file{AGENTS.md},
\nClaudeOnly{} only \file{CLAUDE.md}, and \nBoth{} both. Of those with both, most point one file at the other:
\file{CLAUDE.md} defers to \file{AGENTS.md} through a link, an import or a pointer in \nDefers{} repositories, the
reverse in \nReverse, and only \nTwoReal{} keep two files with content of their own. New adopters have moved from
starting with \file{CLAUDE.md} (\pctClaudeFirstEarly{} from January to August 2025) to starting with
\file{AGENTS.md} (\cref{fig:adoption}B; \pctClaudeFirstLate{} \file{CLAUDE.md}-first since May 2026). Renames,
which path history misses, explain \pctRenamedFromClaude{} of the \file{AGENTS.md}-first cases (95\% CI
\ciRenamed), too few to change the trend. This confirms at population scale the convergence that
\citet{galster2026harness} found in cross-file references (\cref{sec:app-filenames}).

\subsection{RQ2: What do agent instructions contain?}\label{sec:rq2}

\begin{takeaway}
Almost every instruction set contains rules. Code conventions, building, agent workflow, documentation and
testing each appear as rule subjects in most repositories. Collaboration rules name concrete requirements,
from test plans to commit formats.
\end{takeaway}

The median instruction set has \medianWords{} words (10th and 90th percentiles \pTenWords{} and
\pNinetyWords). \pctReposRule\% contain at least one rule (95\% CI \ciReposRule), and on average a repository's
statements are \meanRuleShare\% rules, \meanDescShare\% descriptions and \meanRefShare\% references, such as
commands and pointers (Sonnet's codes, \cref{eq:share}). These are model-defined categories: human agreement is
\pctHumanDescRule{} for description versus rule (references excluded) and \pctHumanArea{} for subject areas
(\cref{sec:review-results}).
The rules govern the code and its artefacts in
\repoRulesCode\% of repositories, the workflow in \repoRulesWorkflow\%, the agent's own behaviour in
\repoRulesAgent\%, and the instruction files themselves in \repoRulesFiles\%.

\begin{figure}[t]
\centering
\includegraphics[width=\linewidth]{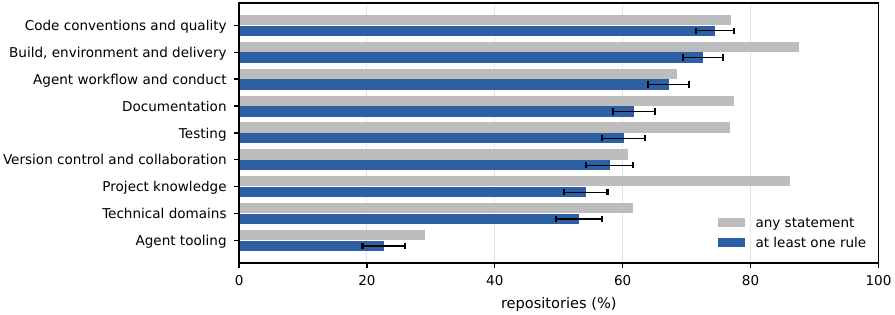}
\caption{Rules most often concern code conventions, building and agent workflow. Share of repositories whose
instruction set has at least one rule (blue, 95\%
bootstrap intervals) and any statement (grey) in each subject area. Sonnet's codes on \nReposEone{} sample repositories, reweighted to
the population.}
\label{fig:content}
\end{figure}

\Cref{fig:content} shows what the rules are about. Code conventions and quality (\areaCode\%), build,
environment and delivery (\areaBuild\%) and agent workflow and conduct (\areaWorkflow\%) lead, followed by documentation and testing.
Project knowledge appears in \areaKnowledgeAny\% of instruction sets, while \areaKnowledgeRule\% contain a rule
on that area. The most common specific rule subjects are verification before declaring
work done (\subjVerif\%), test writing (\subjTestWriting\%) and the instruction files themselves
(\subjAgentFiles\%), each in under half of repositories and with overlapping intervals.

\paragraph{What the rules require.} For four collaboration subjects, rules name concrete parts
(\cref{tab:elements}; Haiku's codes, uncorrected). Rules on pull-request descriptions most often require a test plan
(\prTestPlan\% of the files with such rules), the project's template and linked issues; rules on commit messages
require the Conventional Commits format (\commitConv\%). Rules on test writing prohibit trivial tests
(\testTrivialPro\%) and weakening tests to make them pass (\testWeakenPro\%), and code-review rules disagree on
style (\reviewStyleReq\% ask reviewers to check it, \reviewStylePro\% not to nitpick it). AI attribution splits
projects: \pctRepoAIPro{} of the repositories with commit-message rules prohibit it (95\% CI \ciRepoAIPro) and
\nRepoAIReq{} require it, while for pull requests \pctRepoPRAIReq{} require disclosure of AI use and
\pctRepoPRAIPro{} prohibit it, mostly in different repositories.

\begin{table}[t]
\centering\small
\caption{Selected elements of rules on pull-request descriptions and commit messages: share of the files with a
rule on the subject that require or prohibit each element (Haiku, full corpus, uncorrected). All categories are in
the appendix.}
\label{tab:elements}
\emph{Pull-request descriptions}\par\smallskip
\begin{tabular}{>{\raggedright\arraybackslash}p{0.42\linewidth}rr}
\toprule
Category & \% requiring & \% prohibiting \\
\midrule
test plan / verification evidence & 27.2 & 4.4 \\
use and complete the PR template & 24.2 & 1.6 \\
linked issues and related PRs & 23.3 & 1.3 \\
title format (type, tag or scope prefix) & 19.3 & 1.7 \\
summary of the change & 17.8 & 0.1 \\
screenshots and visual evidence & 15.6 & 1.1 \\
motivation and problem & 12.9 & 0.2 \\
concision, length limits, excluded noise & 9.3 & 4.8 \\
AI-use disclosure and attribution & 5.7 & 1.9 \\
\bottomrule
\end{tabular}
\par\medskip
\emph{Commit messages}\par\smallskip
\begin{tabular}{>{\raggedright\arraybackslash}p{0.42\linewidth}rr}
\toprule
Category & \% requiring & \% prohibiting \\
\midrule
Conventional Commits / structured header & 36.7 & 1.8 \\
allowed commit types & 23.5 & 0.6 \\
subject mood, case, punctuation & 15.1 & 1.3 \\
scope prefix & 14.9 & 0.8 \\
issue / PR reference & 11.4 & 1.7 \\
subject length limit & 10.0 & 0.0 \\
sign-off and human authorship & 5.9 & 1.9 \\
AI / agent attribution & 3.5 & 7.4 \\
\bottomrule
\end{tabular}

\end{table}

\subsection{RQ3: How are agent instructions expressed?}\label{sec:rq3}

\begin{takeaway}
Instruction sets speak to an unnamed agent in plain imperatives. Mandatory rules and prohibitions each appear
in most instruction sets, but unmarked imperatives are more common. Most directives give no reason.
\end{takeaway}

\begin{figure}[t]
\centering
\includegraphics[width=\linewidth]{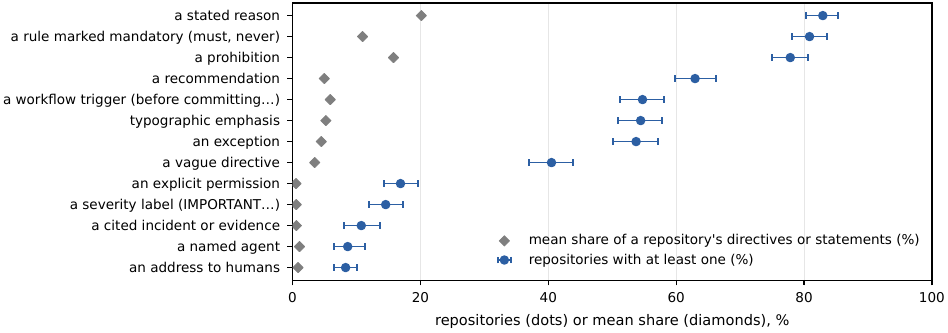}
\caption{Most instruction sets contain marked rules, prohibitions and reasons somewhere, but each is a small share
of their directives. Share of repositories with at least one directive of each kind (dots, 95\%
bootstrap intervals) and its mean share of a repository's directives or statements (diamonds). Sonnet's codes on
the sample, reweighted.}
\label{fig:style}
\end{figure}

In \cref{fig:style}, \pctReposMandatory{} of instruction sets mark at least one
rule as mandatory and \pctReposProhibit{} prohibit something, yet prohibitions are on average \meanProhibShare{} of
a repository's directives, and unmarked imperatives are the most common form. \pctReposReason{} give a reason at
least once, but \pctDirectivesNoReason{} of directives carry none, and only \pctReposIncident{} of repositories cite
an incident or evidence. Rules are often tied to a moment in the workflow (\pctReposLifecycle) or state an
exception (\pctReposException). Almost every directive addresses the agent generically; \pctNameAgent{} of
repositories name a specific agent somewhere, and \pctAddressHumans{} address human contributors. Vague directives
appear in \pctReposVague{} of repositories but are only \pctVagueShare{} of directives.

\subsection{RQ4: Where does the content come from?}\label{sec:rq4}

\begin{takeaway}
One in three instruction sets shows a trace of a generator, a tool, another project or an AI-marked creation, and
four in five are edited after creation.
\end{takeaway}

\paragraph{Traces of origin.} For each of the \nDocuments{} instruction sets we combined the traces left in the
text with the history of the files that carry it (\cref{tab:origins} in the appendix). \pctTextTrace{} carry a
trace in the text itself. Of the \nOriginHist{} instruction sets with a file history of their own, \pctAnyTrace{}
carry a trace in the text or were created in an AI-marked commit. These are traces of reuse or assistance,
not a classification of whole files as human- or machine-written.

\paragraph{Generator signatures.} \nInitSig{} instruction sets (\pctInitSig) match a signature associated with an
\texttt{/init} command. Most common is the opening line of \texttt{claude /init}, in \nClaudeInit{} (\pctClaudeInit); the heading
set of Codex's \texttt{/init} appears in \nCodexInit. The share falls with popularity, from \pctGenLow{} of
instruction sets in the lowest star tier to \pctGenTop{} in the highest. Detection is uneven across tools:
\texttt{claude /init} leaves a verbatim line, Codex's \texttt{/init} only a set of headings, and other tools no
signature at all, so these shares mainly measure Claude Code's generator.

\paragraph{Tools that write into agent files.} \nToolBlock{} instruction sets (\pctToolInserted) contain a
block that a framework or tool writes into the file itself. The Next.js development server inserts
one in \nNextjsBlock{}; it warns the model that the framework's APIs may differ from its training data. Turborepo,
Nx, Laravel Boost and the beads issue tracker write similar blocks.

\paragraph{Shared text across projects.} \pctSharedText{} of instruction sets share a paragraph verbatim with the files
of at least two other owners. About half of these also carry generator text or a tool block; the remaining
\nSharedOnly{} instruction sets (\pctSharedOnly) share text without either signature, including a common set of
behavioural guidelines for coding agents (\nCopiedRuleSet{} instruction sets). Shared text may come from another
project or a common source. Whole-file similarity is less frequent:
\pctCrossDup{} of instruction sets are near-duplicates of another owner's file.

\paragraph{AI-marked creation.} \pctOriginAICreated{} of the instruction sets were created in an AI-marked commit, and
\pctOriginAnyAI{} have at least one AI-marked commit. An AI-marked edit after creation appears in
\pctOriginAIEdit{}. Counted by file, the
creating commit is AI-marked for \pctAICreatedAgents{} of \file{AGENTS.md} and \pctAICreatedClaude{} of
\file{CLAUDE.md} files, and \pctAICommits{} of the \nAgentFileCommits{} commits that touched agent files are
AI-marked. These markers do not establish which text the agent wrote, and they mostly capture one tool: among adopters with any AI-marked
commit in their latest \commitSample{} commits, \pctClaudeOfAIUsers{} have a Claude marker, and tools differ in whether they
add a trailer by default.

\paragraph{Maintenance.} \pctOriginChanged{} of instruction sets were edited after creation. Those with generator
signatures were edited about as often as those without (\pctGenChanged{} versus \pctNoGenChanged{}); neither
share is adjusted for age. By file, \pctAgentsMulti{} of \file{AGENTS.md} and \pctClaudeMulti{} of
\file{CLAUDE.md} files have more than one commit. By comparison, \citet{mohsenimofidi2026contexteng} found that
half of their \file{AGENTS.md} files had never changed.

\subsection{RQ5: What does content vary with?}\label{sec:rq5}

\begin{takeaway}
The prespecified tests associate popularity with pull-request and AI-policy rules, and compiled languages with
build instructions. Exploratory analyses link activity to process-oriented content and find LLM and agent
projects over-represented among adopters.
\end{takeaway}

\paragraph{Prespecified tests.} Of the six hypotheses specified before fitting the association models (\cref{tab:stageD}), three
are supported: more-starred repositories more often have rules on pull requests or AI contribution policy (H6;
OR \hSixOR{} per standard deviation of log stars), compiled-language repositories more often state how to build and
run the project (H3; OR \hThreeOR), and \file{CLAUDE.md} is more descriptive than \file{AGENTS.md} in the same
repository (H5). More merged pull requests go with more workflow rules and fewer descriptions, but more human
authors go with slightly fewer workflow rules, so H2 is mixed. The AI-share (H1) and emphasis (H4) hypotheses are
not supported, with small significant effects against the prediction (IRR \hOneAgentIRR{} and OR \hFourOR). H3, H6
and the pull-request part of H2 hold on the Sonnet sample and when either of the two largest languages is left
out. The paired file comparison was informed by earlier pooled differences; the percentage-point estimate and
other post hoc checks are distinguished in \cref{sec:app-deviations}.

\begin{table}[t]
\centering\small
\caption{Prespecified hypotheses: verdicts and principal effects. Ratios are per standard deviation of continuous
predictors; H3 compares compiled with other languages, and H5 is a paired percentage-point difference.
Benjamini--Hochberg correction covers \nTestsD{} tests. \nReposD{} repositories; H5 uses \nPairsD{} pairs.}
\label{tab:stageD}
\begin{tabularx}{\linewidth}{@{}>{\raggedright\arraybackslash}p{0.30\linewidth}>{\raggedright\arraybackslash}p{0.15\linewidth}>{\raggedright\arraybackslash}X@{}}
\toprule
Hypothesis & Verdict & Effect (95\% CI) \\
\midrule
H1 AI-marked share: more agent rules, permission/escalation rules and prohibitions & not supported & IRR 0.977 (0.959--0.996), 0.992 (0.963--1.021), 0.981 (0.965--0.997) \\
H2 Collaboration: more workflow rules, fewer descriptions & mixed (2 of 4) & merged PRs: workflow IRR 1.027 (1.009--1.046), descriptions 0.876 (0.858--0.894); human authors: workflow 0.975 (0.959--0.992, opposite), descriptions 0.993 (n.s.) \\
H3 Compiled languages: more build/run statements & supported & OR 1.68 (1.48--1.91) \\
H4 Emphasis: more rules on agent failure modes & not supported & OR 0.93 (0.88--0.99) \\
H5 \file{CLAUDE.md} more descriptive than \file{AGENTS.md}, same repo & supported & +6.3 pp (4.2--8.4); Wilcoxon p = 1.3e-7 \\
H6 Stars: more pull-request or AI-policy rules & supported & OR 1.36 (1.29--1.44) \\
\bottomrule
\end{tabularx}

\end{table}

\begin{figure}[t]
\centering
\includegraphics[width=\linewidth]{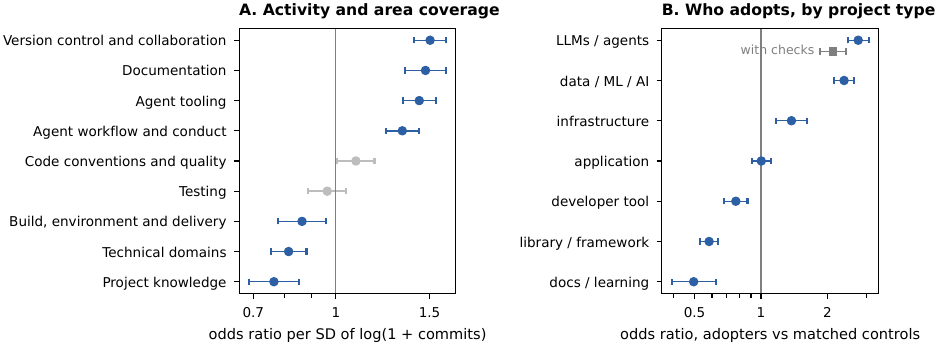}
\caption{Activity goes with process- and agent-oriented content, and AI projects are over-represented among
adopters. \textbf{(A)} Exploratory models: odds ratio of covering each area per standard deviation of
$\log(1 + \text{commits in the past year})$, with other predictors held fixed (blue: significant after
Benjamini--Hochberg, grey: not; Haiku's codes,
\nReposE{} instruction sets). \textbf{(B)} Adopters against \nControlsCompare{} matched repositories without agent
files, by project type, within star tier and language; the square adds controls for age and activity and
excludes agent-tooling projects. Project types are model-classified and indicative.}
\label{fig:variation}
\end{figure}

\paragraph{Exploratory associations.} In the exploratory models, \nSigE{} of \nTestsE{} tests pass the
false-discovery correction (\cref{tab:e-predictors}). Activity is associated with the largest number of tested
content outcomes (\cref{fig:variation}A): per standard deviation of
log-transformed commits in the past year, the odds of
covering version control (OR \orVcCommits), documentation, agent tooling and agent workflow are higher, and those
of covering project knowledge lower (\orKnowledgeCommits). Older repositories more often cover testing
(OR \orTestingAge{} per standard deviation of age).

\paragraph{Who adopts.} We classified project type using Haiku on repository metadata. It agrees with Sonnet
($\kappa$ \kappaTypeSonnet) but much less with the human reviewer's blind check ($\kappa$ \kappaTypePI, 95\% CI
\ciKappaTypePI), so type contrasts are indicative; AI focus is more reliable ($\kappa$ \kappaAIPI). Projects about
LLMs, generative AI or agents are \pctAIAdopters{} of adopters against \pctAIControls{} of matched controls, an
odds ratio of \orAIFocus{} within star tier and language (\cref{fig:variation}B). In a post hoc check the association shrinks to
\orAIBoth{} (95\% CI \ciAIBoth) when age and activity are held fixed and agent-tooling projects are excluded, but
remains positive. Libraries (OR \orLibrary) and documentation repositories (\orDocs) are under-represented among adopters.
Project-type frequencies and matched comparisons are in \cref{tab:e-types,tab:e-control}.

\FloatBarrier
\section{Threats to validity}\label{sec:threats}

\paragraph{Construct validity.} We measure the presence and content of root \file{AGENTS.md} and \file{CLAUDE.md}
files, not agent use or the instructions agents receive at run time. Nested agent files are excluded; other
formats are counted only in RQ1. AI-marked commits capture only part of AI-assisted work, depend on whether a tool adds a trailer, and
are depressed in projects that ban trailers. Project type is classified from self-written metadata and agrees only
moderately with a human reading.

\paragraph{Measurement.} Content categories were discovered and applied by language models. They are reproducible
across models but diverge from the human reviewer's reading, including on broad subject areas. The review covers
\nHumanCoded{} statements with fully matching source spans, from validation files; it excludes languages the
reviewer could not read and stopped early. Its acceptance rates describe this sample, not corpus-wide accuracy.
Topic and element-stance results are model-defined. Annotation error can bias their associations with project
characteristics. Generator signatures and shared text are evidence of possible reuse or assistance, with unknown
false-positive rates; AI-marked commits do not identify which text an agent wrote.
The models analysed files of their vendor's format; they may treat Claude-specific conventions differently from
others, which we could not test directly.

\paragraph{Internal validity.} History covers only files that survive at the snapshot, and path history does not
follow renames; a rename check bounds the second effect. Claude Code began reading \file{AGENTS.md} four weeks
before the snapshot, so recent shares may reflect that change. All correlations are cross-sectional, and activity
counts may include commits made by agents.

\paragraph{External validity and inference.} The population is popular, recently active repositories in thirteen
languages on one snapshot date; stars can be gamed and differ by language \citep{he2026fakestars,borges2016popularity}.
With almost six thousand repositories, small effects in the exploratory models become significant; the results are
exploratory leads, controlled for false discoveries only within their family.

\section{Discussion}\label{sec:discussion}

\paragraph{Instructions as onboarding with expectations.} Agent instructions read like a compact onboarding
document: what the project is, which commands to run, and how to work. Verification before declaring work done,
test writing, the instruction files themselves and code design are among the most common specific rule subjects,
yet each appears in under half of repositories. Projects share broad concerns more than individual requirements.
Project knowledge appears in \areaKnowledgeAny\% of instruction sets, with rules on that area in
\areaKnowledgeRule\%. The files give an agent starting a task both project context and expectations for its work.

\paragraph{Concrete, but mostly unexplained.} Rules on collaboration artefacts name specific requirements, such as a
test plan, linked issues or a commit format; in agent skills, a preprint finds that concrete instructions of this
kind carry most of the measured gains \citep{wang2026skillevolution}. Yet most directives give no reason, few cite an
incident, and marked obligations and prohibitions are a minority of directives. Whether reasons or emphasis change
what agents do remains open.

\paragraph{Reused, assisted and maintained.} A third of instruction sets carry a generator-associated signature,
tool-inserted block, shared text or AI-marked creation commit. Frameworks even write into these files themselves.
Four in five sets are edited after creation. Studies that compare developer-written with generated files
\citep{gloaguen2026evaluating} therefore face a practical complication: a file can combine tool-supplied text,
reused guidelines and later project edits.

\paragraph{One instruction set.} Most repositories with both file names link one to the other, usually with
\file{AGENTS.md} as the substantive file and \file{CLAUDE.md} as a pointer. Counting each file as an independent
observation would duplicate many repositories; resolving links and imports reveals what the agent is asked to read.

\paragraph{AI attribution has no shared norm.} Projects that regulate AI attribution in commit messages more often
prohibit it than require it, while rules on pull-request descriptions more often require disclosure, mostly in other
projects. Trailer-based measures of AI use inherit this disagreement: projects with equal AI use can leave
different traces if one prohibits attribution and the other permits it.

\paragraph{What content varies with.} In exploratory models, activity is associated with more of the tested
content outcomes than other project characteristics: active projects more often cover version control, documentation and how the agent
should work. LLM and agent projects remain over-represented among adopters after age, activity and agent-tooling
projects are accounted for. This fits early adoption by the community closest to the tools, but adoption may also
shape how projects describe themselves. The cross-sectional data cannot establish the direction of these
associations, and activity counts may themselves include agent commits.

\paragraph{A method lesson.} The models agreed with one another much more closely than with the human reviewer.
Agreement among models alone therefore offers limited assurance that their categories match a human
interpretation. The difference is especially important for topics: even the broad subject areas had low human
agreement, while agreement was higher for the distinction between descriptions and rules.

\paragraph{Future work.} Whether repositories follow their own instructions, for example their AI-attribution rules,
can be measured from their commits, and the exploratory associations need confirmatory tests on fresh data.

\section{Conclusion}\label{sec:conclusion}

One in four repositories in our census of popular, active GitHub projects in thirteen languages has root
instructions for coding agents. These files put project knowledge beside concrete expectations about code,
testing and collaboration. They also expose unsettled norms: projects disagree about AI attribution, and most
directives offer no explanation. Generator signatures, shared text and tool-inserted blocks reveal the role of
reusable text, while file histories show that projects keep editing the instructions. Exploratory analyses associate
activity with process-oriented content and find LLM and agent projects over-represented among adopters.
The census and released annotations provide a basis for studying how this practice develops, and whether agents
follow the expectations projects write down.

\section*{AI-use statement}

We used Claude Haiku, Sonnet and Opus 5.5 for segmentation, annotation, project classification and taxonomy
development (\cref{sec:stage-c,sec:llm-use}). Claude Code assisted with analysis code and drafting;
OpenAI Codex and GPT models assisted with revision and review. The author designed the study, made the analytic decisions,
reviewed and updated the source code, edited the text, and is responsible for the content.

\section*{Data availability}

We release, for every repository in the census, its identifier and adoption status, and for every analysed file
its repository, path and the commit it was read at, so that the corpus can be reconstructed. We also release the
statement-level annotations, taxonomies and aggregate outputs, without redistributing third-party file texts.
The dataset is available on
\href{https://huggingface.co/datasets/pzarzycki/agents-claude-md-github-census-2026}{Hugging Face}.

\bibliographystyle{plainnat}
\bibliography{references}

@inbook{chatlatanagulchai2025manifests, title={{On the Use of Agentic Coding Manifests: An Empirical Study of Claude Code}}, ISBN={9783032120892}, ISSN={1611-3349}, url={http://dx.doi.org/10.1007/978-3-032-12089-2_40}, DOI={10.1007/978-3-032-12089-2_40}, booktitle={Product-Focused Software Process Improvement}, publisher={Springer Nature Switzerland}, author={Chatlatanagulchai, Worawalan and Thonglek, Kundjanasith and Reid, Brittany and Kashiwa, Yutaro and Leelaprute, Pattara and Rungsawang, Arnon and Manaskasemsak, Bundit and Iida, Hajimu}, year={2025}, month=nov, pages={543--551} }

@article{chatlatanagulchai2026agentreadmes, title={{Agent READMEs: An Empirical Study of Context Files for Agentic Coding}}, ISSN={1557-7392}, url={http://dx.doi.org/10.1145/3840295}, DOI={10.1145/3840295}, journal={ACM Transactions on Software Engineering and Methodology}, publisher={Association for Computing Machinery (ACM)}, author={Chatlatanagulchai, Worawalan and Li, Hao and Kashiwa, Yutaro and Reid, Brittany and Thonglek, Kundjanasith and Leelaprute, Pattara and Rungsawang, Arnon and Manaskasemsak, Bundit and Adams, Bram and Hassan, Ahmed E. and Iida, Hajimu}, year={2026}, month=sep }

@inproceedings{mohsenimofidi2026contexteng, series={MSR '26}, title={{Context Engineering for AI Agents in Open-Source Software}}, url={http://dx.doi.org/10.1145/3793302.3793350}, DOI={10.1145/3793302.3793350}, booktitle={Proceedings of the 23rd International Conference on Mining Software Repositories}, publisher={ACM}, author={Mohsenimofidi, Seyedmoein and Galster, Matthias and Treude, Christoph and Baltes, Sebastian}, year={2026}, month=apr, pages={194--198}, collection={MSR '26} }

@inproceedings{santos2026decoding, series={AGENT '26}, title={{Decoding the Configuration of AI Coding Agents: Insights from Claude Code Projects}}, url={http://dx.doi.org/10.1145/3786167.3788412}, DOI={10.1145/3786167.3788412}, booktitle={Proceedings of the 2026 International Workshop on Agentic Engineering}, publisher={ACM}, author={Santos, Hélio Victor Flexa dos and Costa, Vitor and Montandon, João Eduardo and Valente, Marco Tulio}, year={2026}, month=apr, pages={63--67}, collection={AGENT '26} }

@inproceedings{jiang2026cursorrules, series={MSR '26}, title={{Beyond the Prompt: An Empirical Study of Cursor Rules}}, url={http://dx.doi.org/10.1145/3793302.3793367}, DOI={10.1145/3793302.3793367}, booktitle={Proceedings of the 23rd International Conference on Mining Software Repositories}, publisher={ACM}, author={Jiang, Shaokang and Nam, Daye}, year={2026}, month=apr, pages={397--409}, collection={MSR '26} }

@inproceedings{sun2026cursorrulesfiles, title={{A Study of Cursorrules Files in GitHub Open Source Projects}}, url={http://dx.doi.org/10.5220/0015072600004088}, DOI={10.5220/0015072600004088}, booktitle={Proceedings of the 21st International Conference on Software Technologies}, publisher={SCITEPRESS - Science and Technology Publications}, author={Sun, Shuang and Akhoundali, Jafar and Kudriavtseva, Arina and Karayalcin, Sengim and Gadyatskaya, Olga}, year={2026}, pages={51--62} }

@inproceedings{galster2026configuring, series={AIware '26}, title={{Configuring Agentic AI Coding Tools: An Exploratory Study}}, url={http://dx.doi.org/10.1145/3805760.3814887}, DOI={10.1145/3805760.3814887}, booktitle={Proceedings of the 3rd ACM International Conference on AI-Powered Software}, publisher={ACM}, author={Galster, Matthias and Mohsenimofidi, Seyedmoein and Lulla, Jai Lal and Abubakar, Muhammad Auwal and Treude, Christoph and Baltes, Sebastian}, year={2026}, month=jul, pages={11--20}, collection={AIware '26} }

@misc{galster2026harness,
      title={{Harness Engineering for Agentic AI Coding Tools: An Exploratory Study}}, 
      author={Matthias Galster and Seyedmoein Mohsenimofidi and Jai Lal Lulla and Muhammad Auwal Abubakar and Christoph Treude and Sebastian Baltes},
      year={2026},
      eprint={2602.14690},
      archivePrefix={arXiv},
      primaryClass={cs.SE},
      url={https://arxiv.org/abs/2602.14690}, 
}

@inproceedings{galster2026dataset, series={AIware '26}, title={{A Dataset of Agentic AI Coding Tool Configurations}}, url={http://dx.doi.org/10.1145/3805760.3814922}, DOI={10.1145/3805760.3814922}, booktitle={Proceedings of the 3rd ACM International Conference on AI-Powered Software}, publisher={ACM}, author={Galster, Matthias and Mohsenimofidi, Seyedmoein and Böhme, Levi and Lulla, Jai Lal and Abubakar, Muhammad Auwal and Treude, Christoph and Baltes, Sebastian}, year={2026}, month=jul, pages={314--322}, collection={AIware '26} }

@misc{santos2026smells,
      title={{Configuration Smells in AGENTS.md Files: Common Mistakes in Configuring Coding Agents}}, 
      author={Helio Victor F. dos Santos and Vitor Costa and Joao Eduardo Montandon and Luciana Lourdes Silva and Marco Tulio Valente},
      year={2026},
      eprint={2606.15828},
      archivePrefix={arXiv},
      primaryClass={cs.SE},
      url={https://arxiv.org/abs/2606.15828}, 
}

@misc{chakrabarti2026growing,
      title={{Why Does CLAUDE.md Keep Growing? Catastrophic Remembering in Agentic Coding}}, 
      author={Kushal Chakrabarti},
      year={2026},
      eprint={2608.11095},
      archivePrefix={arXiv},
      primaryClass={cs.AI},
      url={https://arxiv.org/abs/2608.11095}, 
}

@misc{cai2026ruletaxonomy,
      title={{Rule Taxonomy and Evolution in AI IDEs: A Mining and Survey Study}}, 
      author={Guangzong Cai and Ruiyin Li and Peng Liang and Zengyang Li and Mojtaba Shahin},
      year={2026},
      eprint={2606.12231},
      archivePrefix={arXiv},
      primaryClass={cs.SE},
      url={https://arxiv.org/abs/2606.12231}, 
}

@inproceedings{treude2026ethics, series={AIware '26}, title={{Operationalizing Ethics for AI Agents: How Developers Encode Values into Repository Context Files}}, url={http://dx.doi.org/10.1145/3805760.3814899}, DOI={10.1145/3805760.3814899}, booktitle={Proceedings of the 3rd ACM International Conference on AI-Powered Software}, publisher={ACM}, author={Treude, Christoph and Baltes, Sebastian and Cheong, Marc}, year={2026}, month=jul, pages={106--109}, collection={AIware '26} }

@misc{diaz2026methharness,
      title={{Methodological Harness in Agentic Software Engineering: An Empirical Study on Mining Software Repositories}}, 
      author={Jessica Díaz and Jorge Pérez and Sergio Gil-Borrás},
      year={2026},
      eprint={2609.32014},
      archivePrefix={arXiv},
      primaryClass={cs.SE},
      url={https://arxiv.org/abs/2609.32014}, 
}

@misc{yan2026donot,
      title={{When "Do Not" Is Not Deny: Security Rules in CLAUDE.md vs Built-In Controls}}, 
      author={Ting Yan},
      year={2026},
      eprint={2608.23550},
      archivePrefix={arXiv},
      primaryClass={cs.HC},
      url={https://arxiv.org/abs/2608.23550}, 
}

@misc{gloaguen2026evaluating,
      title={{Evaluating AGENTS.md: Are Repository-Level Context Files Helpful for Coding Agents?}}, 
      author={Thibaud Gloaguen and Niels Mündler-Sasahara and Mark Niklas Müller and Veselin Raychev and Martin Vechev},
      year={2026},
      eprint={2602.11988},
      archivePrefix={arXiv},
      primaryClass={cs.SE},
      url={https://arxiv.org/abs/2602.11988}, 
}

@misc{lulla2026efficiency,
      title={{On the Impact of AGENTS.md Files on the Efficiency of AI Coding Agents}}, 
      author={Jai Lal Lulla and Seyedmoein Mohsenimofidi and Matthias Galster and Jie M. Zhang and Sebastian Baltes and Christoph Treude},
      year={2026},
      eprint={2601.20404},
      archivePrefix={arXiv},
      primaryClass={cs.SE},
      url={https://arxiv.org/abs/2601.20404}, 
}

@misc{khatri2026ablation,
      title={{Do Context Files Help Coding Agents? A Two-Agent Ablation Study on Real Repositories}}, 
      author={Prakhar Khatri},
      year={2026},
      eprint={2607.27250},
      archivePrefix={arXiv},
      primaryClass={cs.SE},
      url={https://arxiv.org/abs/2607.27250}, 
}

@inproceedings{arabat2026instructions, series={MSR '26}, title={{Toward Instructions-as-Code: Understanding the Impact of Instruction Files on Agentic Pull Requests}}, url={http://dx.doi.org/10.1145/3793302.3793601}, DOI={10.1145/3793302.3793601}, booktitle={Proceedings of the 23rd International Conference on Mining Software Repositories}, publisher={ACM}, author={Arabat, Ali and Sayagh, Mohammed}, year={2026}, month=apr, pages={969--973}, collection={MSR '26} }

@misc{zhang2026guardrails,
      title={{Guardrails Beat Guidance: A Large-Scale Study of Rules, Skills, and Persistent Configuration for Coding Agents}}, 
      author={Xing Zhang and Guanghui Wang and Yanwei Cui and Wei Qiu and Ziyuan Li and Bing Zhu and Peiyang He},
      year={2026},
      eprint={2604.11088},
      archivePrefix={arXiv},
      primaryClass={cs.AI},
      url={https://arxiv.org/abs/2604.11088}, 
}

@misc{mcmillan2026adherence,
      title={{Instruction Adherence in Coding Agent Configuration Files: A Factorial Study of Four File-Structure Variables}}, 
      author={Damon McMillan},
      year={2026},
      eprint={2605.10039},
      archivePrefix={arXiv},
      primaryClass={cs.SE},
      url={https://arxiv.org/abs/2605.10039}, 
}

@misc{wang2026skillevolution,
      title={{Agent Skill Evolution: How Revisions Affect Coding Agents}}, 
      author={Jiajie Wang and Yutong Zhao and Tianlin Li and Huashan Chen and Jinfu Chen and Kebin Peng and Sen He},
      year={2026},
      eprint={2610.04832},
      archivePrefix={arXiv},
      primaryClass={cs.SE},
      url={https://arxiv.org/abs/2610.04832}, 
}

@inproceedings{li2026aidev, series={MSR '26}, title={{AIDev: Studying AI Coding Agents on GitHub}}, url={http://dx.doi.org/10.1145/3793302.3797249}, DOI={10.1145/3793302.3797249}, booktitle={Proceedings of the 23rd International Conference on Mining Software Repositories}, publisher={ACM}, author={Li, Hao and Zhang, Haoxiang and Hassan, Ahmed E.}, year={2026}, month=apr, pages={1029--1033}, collection={MSR '26} }

@article{robbes2026agenticmuch, title={{Agentic Much? Adoption of Coding Agents on GitHub}}, ISSN={1557-7392}, url={http://dx.doi.org/10.1145/3822180}, DOI={10.1145/3822180}, journal={ACM Transactions on Software Engineering and Methodology}, publisher={Association for Computing Machinery (ACM)}, author={Robbes, Romain and Matricon, Théo and Degueule, Thomas and Hora, Andre and Zacchiroli, Stefano}, year={2026}, month=jun }

@misc{robbes2026perils,
      title={{Promises, Perils, and (Timely) Heuristics for Mining Coding Agent Activity}}, 
      author={Romain Robbes and Théo Matricon and Thomas Degueule and Andre Hora and Stefano Zacchiroli},
      year={2026},
      eprint={2601.18345},
      archivePrefix={arXiv},
      primaryClass={cs.SE},
      url={https://arxiv.org/abs/2601.18345}, 
}

@misc{khosravani2026census,
      title={{Detecting AI Coding Agents in Open Source: A Validated Multi-Method Census of 180 Million Repositories}}, 
      author={Arsham Khosravani and Audris Mockus},
      year={2026},
      eprint={2606.24429},
      archivePrefix={arXiv},
      primaryClass={cs.SE},
      url={https://arxiv.org/abs/2606.24429}, 
}

@misc{yu2026simpson,
      title={{Beyond Simpson's Paradox: A Cascade of Confounders in AI Agent Pull-Request Co-Authorship}}, 
      author={Haoran Yu and Xiaochong Jiang and Lifei Liu and Su Wang and Pin Qian and Yihang Chen},
      year={2026},
      eprint={2606.22711},
      archivePrefix={arXiv},
      primaryClass={cs.SE},
      url={https://arxiv.org/abs/2606.22711}, 
}

@misc{gao2026autopilot,
      title={{On Autopilot? An Empirical Study of Human-AI Teaming and Review Practices in Open Source}}, 
      author={Haoyu Gao and Peerachai Banyongrakkul and Hao Guan and Mansooreh Zahedi and Christoph Treude},
      year={2026},
      eprint={2601.13754},
      archivePrefix={arXiv},
      primaryClass={cs.SE},
      url={https://arxiv.org/abs/2601.13754}, 
}

@misc{hora2026aipolicy,
      title={{AI Policy, Disclosure, and Human in the Loop: How Are Contribution Guidelines Adapting to GenAI?}}, 
      author={Andre Hora and Romain Robbes},
      year={2026},
      eprint={2605.16706},
      archivePrefix={arXiv},
      primaryClass={cs.SE},
      url={https://arxiv.org/abs/2605.16706}, 
}

@misc{chen2026visible,
      title={{Making AI Visible, Not Vanished: How AI Policies Reshape Developer Experience on GitHub}}, 
      author={Yunqi Chen and Thomas Zimmermann and Bianca Trinkenreich},
      year={2026},
      eprint={2608.03329},
      archivePrefix={arXiv},
      primaryClass={cs.SE},
      url={https://arxiv.org/abs/2608.03329}, 
}

@article{xiao2026selfadmitted, title={{Self-Admitted GenAI Usage in Open-Source Software}}, volume={52}, ISSN={2326-3881}, url={http://dx.doi.org/10.1109/TSE.2026.3681886}, DOI={10.1109/tse.2026.3681886}, number={6}, journal={IEEE Transactions on Software Engineering}, publisher={Institute of Electrical and Electronics Engineers (IEEE)}, author={Xiao, Tao and Fan, Youmei and Calefato, Fabio and Treude, Christoph and Kula, Raula Gaikovina and Hata, Hideaki and Baltes, Sebastian}, year={2026}, month=jun, pages={1891--1910} }

@article{tufano2026selfadmitted, title={{Developers and generative AI: A study of self-admitted usage in open source projects}}, volume={31}, ISSN={1573-7616}, url={http://dx.doi.org/10.1007/s10664-026-10848-w}, DOI={10.1007/s10664-026-10848-w}, number={4}, journal={Empirical Software Engineering}, publisher={Springer Science and Business Media LLC}, author={Tufano, Rosalia and Pepe, Federica and Zampetti, Fiorella and Mastropaolo, Antonio and Dabi{\'c}, Ozren and Di Penta, Massimiliano and Bavota, Gabriele}, year={2026}, month=apr }

@inproceedings{borges2016popularity, title={{Understanding the Factors That Impact the Popularity of GitHub Repositories}}, url={http://dx.doi.org/10.1109/ICSME.2016.31}, DOI={10.1109/icsme.2016.31}, booktitle={2016 IEEE International Conference on Software Maintenance and Evolution (ICSME)}, publisher={IEEE}, author={Borges, Hudson and Hora, Andre and Valente, Marco Tulio}, year={2016}, month=oct, pages={334--344} }

@inproceedings{he2026fakestars, series={ICSE '26}, title={{Six Million (Suspected) Fake Stars on GitHub: A Growing Spiral of Popularity Contests, Spams, and Malware}}, url={http://dx.doi.org/10.1145/3744916.3764531}, DOI={10.1145/3744916.3764531}, booktitle={Proceedings of the 2026 IEEE/ACM 48th International Conference on Software Engineering}, publisher={ACM}, author={He, Hao and Yang, Haoqin and Burckhardt, Philipp and Kapravelos, Alexandros and Vasilescu, Bogdan and Kästner, Christian}, year={2026}, month=apr, pages={1469--1481}, collection={ICSE '26} }

@article{prana2019readme, title={{Categorizing the Content of GitHub README Files}}, volume={24}, ISSN={1573-7616}, url={http://dx.doi.org/10.1007/s10664-018-9660-3}, DOI={10.1007/s10664-018-9660-3}, number={3}, journal={Empirical Software Engineering}, publisher={Springer Science and Business Media LLC}, author={Prana, Gede Artha Azriadi and Treude, Christoph and Thung, Ferdian and Atapattu, Thushari and Lo, David}, year={2018}, month=oct, pages={1296--1327} }

@inproceedings{elazhary2019guidelines, title={{Do as I Do, Not as I Say: Do Contribution Guidelines Match the GitHub Contribution Process?}}, url={http://dx.doi.org/10.1109/ICSME.2019.00043}, DOI={10.1109/icsme.2019.00043}, booktitle={2019 IEEE International Conference on Software Maintenance and Evolution (ICSME)}, publisher={IEEE}, author={Elazhary, Omar and Storey, Margaret-Anne and Ernst, Neil and Zaidman, Andy}, year={2019}, month=sep, pages={286--290} }

@inproceedings{tourani2017coc, title={{Code of conduct in open source projects}}, url={http://dx.doi.org/10.1109/SANER.2017.7884606}, DOI={10.1109/saner.2017.7884606}, booktitle={2017 IEEE 24th International Conference on Software Analysis, Evolution and Reengineering (SANER)}, publisher={IEEE}, author={Tourani, Parastou and Adams, Bram and Serebrenik, Alexander}, year={2017}, month=feb, pages={24--33} }

@inproceedings{hata2021metamaintenance, title={{Same File, Different Changes: The Potential of Meta-Maintenance on GitHub}}, url={http://dx.doi.org/10.1109/ICSE43902.2021.00076}, DOI={10.1109/icse43902.2021.00076}, booktitle={2021 IEEE/ACM 43rd International Conference on Software Engineering (ICSE)}, publisher={IEEE}, author={Hata, Hideaki and Kula, Raula Gaikovina and Ishio, Takashi and Treude, Christoph}, year={2021}, month=may, pages={773--784} }

@inproceedings{pister2024promptset, series={LLM4Code '24}, title={{PromptSet: A Programmer's Prompting Dataset}}, url={http://dx.doi.org/10.1145/3643795.3648395}, DOI={10.1145/3643795.3648395}, booktitle={Proceedings of the 1st International Workshop on Large Language Models for Code}, publisher={ACM}, author={Pister, Kaiser and Paul, Dhruba Jyoti and Joshi, Ishan and Brophy, Patrick}, year={2024}, month=apr, pages={62--69}, collection={LLM4Code '24} }

@misc{cardoen2026ghagentfiles,
  title  = {{{GHAgentFiles}: A dataset of coding agent file histories in {GitHub} repositories}},
  author = {Cardoen, Guillaume and Mens, Tom and Decan, Alexandre},
  year   = {2026},
  howpublished = {Author version, ORBi UMONS},
  url    = {https://orbi.umons.ac.be/bitstream/20.500.12907/56423/1/main.pdf},
}

@misc{agentsmd2026,
  title  = {{{AGENTS.md}: A simple, open format for guiding coding agents}},
  author = {{AGENTS.md}},
  key    = {AGENTS.md},
  year   = {2026},
  url    = {https://agents.md/},
  note   = {Accessed 2026-10-16}
}

@misc{anthropic2026memory,
  title  = {{How {Claude} remembers your project ({Claude Code} documentation)}},
  author = {{Anthropic}},
  year   = {2026},
  url    = {https://code.claude.com/docs/en/memory},
  note   = {Accessed 2026-10-16}
}

@misc{anthropic2026changelog,
  title  = {{{Claude Code} {CHANGELOG}, version 2.1.277: ``Added {AGENTS.md} support''}},
  author = {{Anthropic}},
  year   = {2026},
  url    = {https://github.com/anthropics/claude-code/blob/main/CHANGELOG.md},
  note   = {v2.1.277 published to npm 2026-09-18; accessed 2026-10-16}
}

@article{rogan1978prevalence, title={{Estimating Prevalence from the Results of a Screening Test}}, author={Rogan, Walter J. and Gladen, Beth}, journal={American Journal of Epidemiology}, volume={107}, number={1}, pages={71--76}, year={1978}, doi={10.1093/oxfordjournals.aje.a112510} }

@article{benjamini1995fdr, title={{Controlling the False Discovery Rate: A Practical and Powerful Approach to Multiple Testing}}, author={Benjamini, Yoav and Hochberg, Yosef}, journal={Journal of the Royal Statistical Society: Series B (Methodological)}, volume={57}, number={1}, pages={289--300}, year={1995}, doi={10.1111/j.2517-6161.1995.tb02031.x} }

@article{cohen1960kappa, title={{A Coefficient of Agreement for Nominal Scales}}, author={Cohen, Jacob}, journal={Educational and Psychological Measurement}, volume={20}, number={1}, pages={37--46}, year={1960}, doi={10.1177/001316446002000104} }

@article{gilardi2023chatgpt, title={{{ChatGPT} Outperforms Crowd Workers for Text-Annotation Tasks}}, author={Gilardi, Fabrizio and Alizadeh, Meysam and Kubli, Ma{\"e}l}, journal={Proceedings of the National Academy of Sciences}, volume={120}, number={30}, pages={e2305016120}, year={2023}, doi={10.1073/pnas.2305016120} }

@article{ziems2024css, title={{Can Large Language Models Transform Computational Social Science?}}, author={Ziems, Caleb and Held, William and Shaikh, Omar and Chen, Jiaao and Zhang, Zhehao and Yang, Diyi}, journal={Computational Linguistics}, volume={50}, number={1}, pages={237--291}, year={2024}, doi={10.1162/coli_a_00502} }

\appendix
\section{Method details and further tables}\label{sec:appendix}

\FloatBarrier
\subsection{Method details}\label{sec:app-methods}

This section gives the full procedure behind \cref{sec:methods}.

\paragraph{Estimators.} The existing-taxonomy comparison corrects the share $q$ of files that the model labels with a type for the model's
sensitivity $\mathrm{Se}$ and specificity $\mathrm{Sp}$ on all \nGold{} reference files, clipped to $[0,1]$
\citep{rogan1978prevalence}:
\begin{equation}\label{eq:rogan}
  \hat{p} \;=\; \frac{q + \mathrm{Sp} - 1}{\mathrm{Se} + \mathrm{Sp} - 1}.
\end{equation}
For statement-level distributions in the full corpus, Haiku's code distribution $\boldsymbol{\pi}_H$ is mapped to
Sonnet's with the confusion matrix $P_{jk} = \Pr(\text{Sonnet} = k \mid \text{Haiku} = j)$ estimated on the overlap,
\begin{equation}\label{eq:correction}
  \hat{\boldsymbol{\pi}}_S \;=\; P^{\top} \boldsymbol{\pi}_H .
\end{equation}
For the presence of a code in a repository, with $q$ the share of repositories where Haiku finds it and $p_1$, $p_0$
the probabilities that Sonnet finds it given that Haiku does or does not, the calibrated share is
$\hat{p} = p_1 q + p_0 (1 - q)$. Each calibration is used for a dimension only where it lowers the error on a
held-out half of the overlap.
\subsubsection{Population and census}\label{sec:app-m-population}

The population is every public GitHub repository that, on the snapshot date of \snapshotDate, met four
criteria:
\begin{itemize}
  \item at least \minStars{} stars;
  \item a push within the preceding \activeDays{} days (on or after \activeSince);
  \item neither a fork nor archived;
  \item one of thirteen primary languages as determined by GitHub Linguist: Python, JavaScript, TypeScript,
        Java, C\#, C++, C, Go, Rust, PHP, Ruby, Kotlin or Swift.
\end{itemize}
We enumerated it with GitHub's search API, splitting the star range until each query fell under the search
result cap. This yielded \nPopulation{} repositories. The frame is therefore what GitHub's search index
returns for these criteria. A push counts on any branch and includes pushes by bots.

\paragraph{Census.} We checked every repository in the frame for an \emph{agent file}: a file named exactly
\file{AGENTS.md} or \file{CLAUDE.md} at the root of the default branch. We call a repository with one an
\emph{adopter}. This measures the presence of a file, not whether or how coding agents are used. Nested files,
case variants and other formats are not agent files under this definition; we count the other formats
separately below.

Studies of agent files have mostly drawn on repositories already associated with coding agents, such as the
AIDev dataset \citep{li2026aidev,chatlatanagulchai2026agentreadmes}. Others screened broad populations under
different frames and file definitions \citep{galster2026dataset,cardoen2026ghagentfiles,khosravani2026census}.
Ours is a census of popular, active repositories that is not conditioned on agent use.

\paragraph{Collection.} All data came through GitHub's GraphQL and REST APIs; no repository was cloned. For
adopters, files were read at the snapshot commit. Symbolic links were resolved to their target's text, and the
direction of each link was recorded. We also collected repository metadata and activity over the
\metricsDays{} days before the snapshot, including the human authors and AI-marked commits among the latest
\commitSample{} commits on the default branch.

A commit is \emph{AI-marked} if its message or author e-mail carries a co-author trailer or signature of a
coding agent (Claude, Codex, Copilot, Cursor, Devin, Aider or Gemini). AI-marked commits are a lower bound on
AI-assisted commits. A trailer exists only when a tool or person adds one, and a project can forbid them
(\cref{sec:results}).

\paragraph{How the two files relate.} Each adopter receives one label from an ordered set of rules; the
first match wins. The rules test, in order:
\begin{enumerate}
  \item whether only one file exists;
  \item whether one file is a symbolic link to the other, and in which direction;
  \item whether both files link to a third file;
  \item whether \file{CLAUDE.md} consists of \file{@AGENTS.md} imports or is a short pointer to
        \file{AGENTS.md} (at most \thinImportChars{} characters);
  \item whether it imports \file{AGENTS.md} and adds its own content;
  \item whether \file{AGENTS.md} defers to \file{CLAUDE.md};
  \item whether the two files are identical copies.
\end{enumerate}
Otherwise the files are separate, split by whether one mentions the other. Varying the character threshold from \thresholdLow{} to \thresholdHigh{} leaves the
ratio of deference between the two files unchanged. It does move the count of repositories with two files of
their own content (\cref{sec:results}).

\paragraph{History.} For each of the \nHistoryFiles{} agent files with a resolvable history, we walked the
commit history of its path on the default branch back from the snapshot commit. We recorded the first and last
commits, the number of commits, and whether the creating commit was AI-marked. Two limits apply:
\begin{itemize}
  \item History covers only files that exist at the snapshot.
  \item Path history does not follow renames. A file renamed from another name appears to start at the rename.
\end{itemize}
Trends over time are therefore trends among surviving files, and an \file{AGENTS.md} renamed from
\file{CLAUDE.md} counts as \file{AGENTS.md}-first.

\paragraph{Other formats.} For every repository in the population we listed the root and the directories
\file{.github/}, \file{.cursor/rules/} and \file{.claude/}, one level deep. This counts the other instruction
formats of current coding agents, such as \file{.github/copilot-instructions.md}, \file{GEMINI.md} and Cursor
rules. Adopters were listed at the snapshot commit; the other repositories were listed at their current
\texttt{HEAD} a few hours later.

\subsubsection{Units of analysis}\label{sec:app-m-units}

The object of study is a repository's \emph{instruction set}: everything its root \file{AGENTS.md} and
\file{CLAUDE.md} say, taken together. Symbolic links and thin imports are resolved to their target, so the
same text is never counted twice. The two file names serve the same purpose, and we compare them only in one
short subsection (\cref{sec:rq1}). Each result states its unit.
\begin{description}
  \item[Repositories] for adoption and for the repository-level analyses of \cref{sec:stage-d,sec:stage-e}.
  \item[Documents] for structural features: the text of one instruction set, \nDocuments{}
        in all. Two files with their own content are joined.
  \item[Files] for the existing-taxonomy and statement annotations: every non-link file with text that does not merely defer to the
        other file (\nFilesB{} files). File codes are aggregated to the instruction set for repository-level
        results.
  \item[Statements] for statement-level analysis: the semantic statements into which each file is segmented
        (\cref{sec:stage-c}).
\end{description}

\subsubsection{Structure and wording}\label{sec:app-m-stage-a}

We computed deterministic features of each document from its Markdown structure: size, headings, lists,
code blocks, links and tables. Wording features use prose only, with code removed. They count:
\begin{itemize}
  \item directive words: prohibitions such as \emph{never} and \emph{do not}, mandates such as \emph{must}
        and \emph{always}, and soft guidance such as \emph{should} and \emph{prefer};
  \item capitalised emphasis;
  \item named tools and named agents;
  \item the signatures of known generators such as \file{claude /init}.
\end{itemize}
Readability is the Flesch reading ease, and near-duplicate documents were found with MinHash.

We used model-assisted checks of detector matches in context and corrected or dropped three detectors.
The counts are lexical, not a semantic classification.

\subsubsection{The prior codebook}\label{sec:app-m-stage-b}

To compare our population with earlier work, we applied the \nTypesB{} instruction types of
\citet{chatlatanagulchai2026agentreadmes} to the \nFilesB{} files. Each file is split into sections at its
first- and second-level headings, and overlong sections are split further. We used Haiku 5.5 to label each
section with the types it contains, and a file's labels are the union of its sections' labels.

We calibrated the prompt against the reference study's own human labels: \nGold{} files with stored text,
split into fixed halves of \nGoldHalf{}. The prompt was revised on one half and is reported on the other.
The model assigns more types than the human coders did, so we correct prevalence for its measured
sensitivity and specificity with the Rogan--Gladen estimator \citep{rogan1978prevalence}.

\subsubsection{Statements, subjects and elements}\label{sec:app-m-stage-c}

We used language models to discover categories and annotate the files, with human review of sampled
statements. The reviewer coded these statements before seeing the model annotations. The results distinguish
acceptance of extracted text from agreement with the models' interpretation.

\paragraph{Discovering the categories.}
\begin{itemize}
  \item \textbf{Open coding.} We used Sonnet in Claude Code to generate preliminary codes in
        \nDiscoveryAgents{} batches covering \nDiscoveryDocs{} documents. The documents were stratified by
        language, star tier and file composition, with purposive outliers added. Each batch returned supporting
        quotations with line numbers.
  \item \textbf{Building the codebook.} We used Opus to consolidate the proposed codes, then tested for missing categories
        in two rounds on \nSaturationDocs{} unseen documents. We revised the codebook, used a further model
        critique to look for inconsistencies, and fixed the version used for annotation.
  \item \textbf{Result.} The codebook has \nDimensions{} dimensions and \nCodebookValues{} values.
\end{itemize}
The central dimension is the \emph{function} of a statement (\nFunctions{} values). For reporting, each
function maps to one answer of two coarse questions: \emph{what is it?} (a description, a reference or a
rule) and, for rules, \emph{what does it govern?} (the code and its artefacts, the workflow, the agent's own
behaviour, or the instruction files). The full codebook is released with the dataset.

\paragraph{Statements.} We used Haiku to segment each file into \emph{statements}: units of meaning that a careful
reader would sum up as one claim or instruction. A lead-in stays with its list, and independent rules under
one heading are separated. Statements are cut by meaning, never by lines or punctuation. For each statement
the model returned:
\begin{itemize}
  \item its verbatim spans;
  \item a self-contained restatement;
  \item a free-text subject;
  \item its \emph{elements}: the parts it lists or requires, such as the items a pull-request description
        must contain.
\end{itemize}
This gave \nStatements{} statements in \nFilesSegmented{} files. In \pctVerbatim{} of them the spans were
found verbatim in the source; this checks that spans are copied text, not that the segmentation is right.

\paragraph{Subjects.} We used Opus to propose a two-level subject taxonomy from \nTaxonomySample{} random
statements. We checked its coverage by assigning \nTaxonomyCheck{} fresh statements using Haiku, and \pctTaxonomyOther{} fell
into \emph{other}. The taxonomy has \nSubjects{} subjects in \nAreas{} areas. It was used as proposed,
without a revision round.

\paragraph{Coding.} Each statement was coded on the codebook's dimensions and its subject, in context: each
request held consecutive statements with their spans and restatements, plus the surrounding text.
\begin{itemize}
  \item \textbf{Corpus annotation.} We used Haiku to code the corpus.
  \item \textbf{Sample annotation.} We used Sonnet on a sample of \nSampleFiles{} files, stratified by file name, star tier and
        language (\nSampleFilesCoded{} coded, \nSampleStatements{} statements).
  \item \textbf{Reliability} was measured on \nValFiles{} validation files (\nValStatements{} statements)
        coded by all three models. Opus and Sonnet agree at Cohen's $\kappa$ \kappaStrongModels{} by
        dimension \citep{cohen1960kappa}. Haiku agrees with Sonnet at $\kappa$ \kappaHaikuSonnet.
  \item \textbf{Intervals.} Statement-level distributions use \nBoot{} cluster-bootstrap resamples over
        files. Headline repository-level estimates instead use inverse-inclusion weights and a bootstrap over
        repositories, as in \cref{eq:share}.
  \item \textbf{Corpus-wide distributions} use Haiku's codes, corrected toward Sonnet's with the confusion
        matrix $P(\text{Sonnet code} \mid \text{Haiku code})$.
        \begin{itemize}
          \item The matrix was fitted on a seeded half of the overlap files and tested on the other half,
                for single-code dimensions.
          \item The correction lowered the error on every such dimension, so it is used throughout, refitted
                on the whole overlap.
          \item The corrected distributions are point estimates without intervals.
        \end{itemize}
\end{itemize}
Haiku coded \nHaikuCoded{} statements. In \nGapFiles{} files the model left some statements out of its
responses, and those statements are excluded.

\paragraph{Elements.} For subjects whose rules name concrete parts, we built element taxonomies.
\begin{itemize}
  \item \textbf{Subjects.} We chose four: pull-request descriptions, commit messages, test
        writing and code review. Each was checked to be one that Haiku identifies with a precision of at
        least \elementPrecision{} against Sonnet on the sample overlap. Two other candidates fell below that
        threshold and were not analysed.
  \item \textbf{Categories.} For each subject, we used Opus to propose categories from \nElementSample{}
        sampled elements of its rule statements.
  \item \textbf{Assignment.} We used Haiku to assign each element a category and its
        \emph{stance}: required, prohibited, or only mentioned. Stance is judged per element, because one
        statement can require some parts and prohibit others.
  \item \textbf{Rules without elements.} A rule statement without elements counts as one element, because
        list items are often segmented as separate statements.
\end{itemize}
Subject selection was checked against Sonnet; element categories and stances are model assignments.

The human-review procedure and agreement results are reported in \cref{sec:review-results}.

\subsubsection{Prespecified hypothesis tests}\label{sec:app-m-stage-d}

After inspecting descriptive results, we committed the hypotheses, outcomes, predictors, models,
multiple-testing correction and robustness checks before fitting the hypothesis-testing models
(commit \texttt{a08c5b7}).

\paragraph{Sample.} The unit is the repository. The sample has \nReposD{} adopters with at least
\minStatementsD{} coded statements, excluding those whose default-branch head predates \activeSince.

\paragraph{Outcomes and predictors.}
\begin{itemize}
  \item \textbf{Outcomes} are counts or indicators built from Haiku's statement codes, uncorrected, per
        repository. Annotation error can bias these associations.
  \item \textbf{Predictors:} the share of AI-marked commits among the latest \commitSample{} commits, human
        authors, merged pull requests, compiled language, rate of emphasis, and stars.
\end{itemize}

\paragraph{Models and tests.}
\begin{itemize}
  \item \textbf{Models.} Counts use negative binomial regression with an offset; indicators use logistic
        regression.
  \item \textbf{Controls.} Every model controls for language, star tier and repository age, except that
        language is dropped where compiled language is the predictor, and star tier where stars are the
        predictor. Models of indicators also control for the number of statements.
  \item \textbf{Inference.} Tests are two-sided, with heteroskedasticity-robust (HC1) standard errors. Effects
        are incidence-rate or odds ratios per standard deviation of continuous predictors; the compiled-language
        indicator compares compiled with other languages. Skewed count predictors use log transformations before standardisation.
  \item \textbf{The paired hypothesis.} One hypothesis compares \file{AGENTS.md} with \file{CLAUDE.md} within
        repositories that keep two files with real content. It uses \nPairsD{} repositories whose two files
        each have at least \minStatementsD{} coded statements, and a Wilcoxon signed-rank test. It was
        prespecified as \emph{informed}, because pooled differences between the two file names had already
        been seen.
  \item \textbf{Multiple testing.} The Benjamini--Hochberg procedure \citep{benjamini1995fdr} controls the
        false discovery rate at $q = 0.05$ over the \nTestsD{} prespecified tests.
  \item \textbf{Verdict.} A hypothesis is \emph{supported} if every one of its tests has the predicted
        direction and an adjusted $p < 0.05$, and \emph{mixed} if only some do.
  \item \textbf{Other analyses.} Exploratory analyses form a separate family and are labelled as such.
        Checks added after the results were seen are labelled post hoc.
  \item \textbf{Robustness.} We repeated the tests:
        \begin{itemize}
          \item on the Sonnet sample with Sonnet's codes;
          \item with an alternative AI measure, namely AI-marked commits on the agent files themselves;
          \item leaving out each of the two largest languages.
        \end{itemize}
\end{itemize}

\subsubsection{Exploratory associations with project characteristics}\label{sec:app-m-stage-e}

We planned the exploratory analyses before computing their results, specifying definitions, outcomes,
models and the test family. These analyses search for associations rather than test directional hypotheses.

\paragraph{Content per instruction set.} Statement codes are pooled over a repository's coded files, counting
identical copies once. For each code we report the share of repositories with at least one statement or rule
with it, and the mean share of a repository's statements. Headline estimates use Sonnet's codes on the
\nReposEone{} sample repositories whose whole instruction set Sonnet coded, weighted by the inverse of their
inclusion probability, because the sample was drawn by file. Intervals come from a weighted cluster bootstrap
over repositories. Census estimates use Haiku's codes, calibrated to Sonnet's where that lowers the error on a
held-out half of the sample.

\paragraph{Project type.} We used Haiku 5.5 to classify adopters and matched controls from repository
metadata alone (name, description, topics, primary language and homepage), never from its agent files, using a
fixed codebook of software types and a separate AI focus. We repeated the classification with Sonnet 5.5 on a random
sample, and the human reviewer checked a stratified sample before seeing model labels, with each repository's README and latest
release available (\cref{sec:rq5}).

\paragraph{Models.} The models and profiles use the \nReposE{} instruction sets with at least
\minStatementsD{} statements and Haiku's raw codes, as in the prespecified tests. Content outcomes (area presence, the mix of kinds, the share of marked and prohibited rules,
and length) are regressed on project type, language, stars, human authors, commits in the past year, age and the
AI-marked share, using the same regression families, with log statement count added for binary outcomes. Adopters are compared with the matched
controls by project type within star tier and language strata. Benjamini--Hochberg controls the false
discovery rate within the exploratory family of \nTestsE{} tests. The models are repeated with active weeks added,
and on the Sonnet sample (\nSonnetE{} repositories) with Sonnet's codes.

\subsubsection{Language models as instruments}\label{sec:app-m-llm-use}

\begin{itemize}
  \item \textbf{Category discovery.} We used Sonnet in Claude Code for preliminary codes and Opus for
        consolidation, revision and critique.
  \item \textbf{Segmentation, coding and taxonomies.} These ran through Amazon Bedrock in October 2026, with
        Anthropic's Claude Haiku 5.5, Sonnet 5.5 and Opus 5.5.
  \item \textbf{Request settings.} Requests used structured output constrained to the codebook's codes, low
        reasoning effort for coding, and prompt caching.
  \item \textbf{Treatment of the corpus.} Every prompt told the model to treat the files as data, not as
        instructions to follow.
  \item \textbf{Refusals.} Some requests for offensive-security repositories were refused. Uncoded files
        and statements were excluded from the relevant estimates.
\end{itemize}

\FloatBarrier
\clearpage
\subsection{Human review and agreement with model annotations}\label{sec:review-results}

\paragraph{Review procedure.} The reviewer was the study's author. The review covered three aspects.
\begin{enumerate}
  \item \textbf{Design decisions.} After reviewing model output, we replaced syntactic units
        with semantic statements, introduced the two coarse questions, and added subjects and elements.
  \item \textbf{Spot checks of the models' reading.} Statements were drawn from the validation files,
        stratified by kind and subject area, and limited to statements whose spans were all verbatim. The
        reviewer coded \nHumanCoded{} of the \nHumanDrawn{} drawn, then stopped because the work took
        longer than planned. The presentation order was random. Five
        statements in languages the reviewer could not read were skipped. For each statement the reviewer checked
        whether it was cut correctly and whether its restatement, elements and free-text subject were right.
  \item \textbf{Coding before seeing model annotations.} Before seeing the model's reading, the
        reviewer answered the two coarse questions, ticked every function label that fitted, and coded binding
        force, polarity and subject. Answers were fixed before the model annotations were revealed. The reviewer
        changed answers afterwards for \nRevisedAfterReveal{} statements, and only the frozen answers are
        scored.
\end{enumerate}

\paragraph{Extraction and agreement.} The reviewer accepted the boundaries of \pctCutOk{} of the \nHumanCoded{} reviewed statements, the restatements
and extracted elements of \pctParaphraseOk, and the free-text subjects of \pctSubjectFits{}
(\cref{tab:supervision}). The blind reading agrees with the models on description versus rule in
\pctHumanDescRule{} of statements when references are excluded, but only at $\kappa$ \kappaHumanKind{} on the three kinds and in
\pctHumanArea{} on the subject area and \pctHumanSubject{} on the subject. The reviewer rarely used
\emph{reference}: a command list with comments read as a
rule, because in this genre one line often describes, refers and instructs at once. The models agree with each
other far better than with this human reading, as in the comparison with the reference study's coders
($\kappa$ \kappaModelsB{} between models against \kappaHumanB{} with humans), a pattern noted for LLM annotation
in other fields \citep{gilardi2023chatgpt,ziems2024css}. These checks support extraction in the reviewed sample;
subject categories and element-category and stance assignments remain model-defined.

\begin{table}[!htbp]
\centering\small
\caption{Human review of \nHumanCoded{} statements. Top: checks of extracted statements. Bottom: agreement of the
blind reading with Haiku, Sonnet and Opus (ranges across models).}
\label{tab:supervision}
\begin{tabular}{>{\raggedright\arraybackslash}p{0.42\linewidth}r}
\toprule
Check & Result (n = 60) \\
\midrule
Statement cut correctly & 55 (92\%; Wilson 95\% CI 82--96\%); 5 ``part missing'' \\
Restatement faithful & 56 (93\%; 84--97\%); 4 ``loses meaning'' \\
Listed parts (elements) right & 56 (93\%); 4 ``some missing'' \\
Free-text subject fits & 57 (95\%) \\
\bottomrule
\end{tabular}
\par\medskip
\begin{tabular}{>{\raggedright\arraybackslash}p{0.42\linewidth}r}
\toprule
Independent blind reading vs the models & Agreement (n = 60) \\
\midrule
Description vs rule (references left out) & 82--83\% (Haiku 40/48, Sonnet 37/45, Opus 39/47) \\
Three kinds & 62--67\% ($\kappa$ 0.33--0.40) \\
Subject area (10) & 43--50\% (26--30 of 60) \\
Subject (57) & 32--37\% (19--22 of 60) \\
\bottomrule
\end{tabular}

\end{table}

\paragraph{Scope of the checks.} A single coder gives no measure of agreement between humans. Wilson
intervals in \cref{tab:supervision} are descriptive summaries of these checks, without adjustment for
stratification or clustering by source file. Acceptance of extracted text does not validate the later
element-category or stance assignments. The three-way kind, functions, binding force and subjects remain
model-defined categories.

\FloatBarrier
\clearpage
\subsection{Adoption by language}\label{sec:app-languages}

\begin{table}[htbp]
\centering\small
\begin{tabular}{lrrrr}
\toprule
Language & Repos & \file{AGENTS.md} & \file{CLAUDE.md} & Either \\
\midrule
TypeScript & 4,048 & 34.4\% & 24.5\% & 40.8\% \\
Rust & 1,706 & 28.1\% & 18.9\% & 32.7\% \\
Go & 2,264 & 24.5\% & 16.6\% & 28.9\% \\
Swift & 474 & 23.4\% & 17.9\% & 28.1\% \\
Kotlin & 611 & 21.1\% & 14.9\% & 26.0\% \\
Java & 1,262 & 20.9\% & 12.4\% & 24.6\% \\
Python & 5,015 & 19.5\% & 14.9\% & 24.3\% \\
C\# & 964 & 18.2\% & 12.0\% & 23.5\% \\
PHP & 804 & 14.6\% & 9.1\% & 18.9\% \\
JavaScript & 2,211 & 13.0\% & 10.1\% & 17.5\% \\
C++ & 1,785 & 14.7\% & 7.7\% & 17.4\% \\
Ruby & 438 & 15.1\% & 8.9\% & 17.1\% \\
C & 1,272 & 10.7\% & 6.3\% & 13.0\% \\
\bottomrule
\end{tabular}

\caption{Repositories with each agent file, by primary language. Census of \nPopulation{} repositories.}
\label{tab:languages}
\end{table}

\FloatBarrier
\subsection{Other formats}\label{sec:app-formats}

\begin{table}[htbp]
\centering\small
\begin{tabular}{>{\raggedright\arraybackslash}p{0.42\linewidth}rrr}
\toprule
Format & Repos & Share & Also has \file{/AGENTS.md} \\
\midrule
\file{/AGENTS.md} & 4,954 & 21.7\% & --- \\
\file{/CLAUDE.md} & 3,440 & 15.1\% & 69.4\% \\
\file{.github/copilot-instructions.md} & 1,024 & 4.5\% & 52.3\% \\
\file{.cursor/rules/} ($\geq$ 1 .md/.mdc) & 347 & 1.5\% & 64.6\% \\
\file{.github/instructions/} & 273 & 1.2\% & 59.0\% \\
\file{/GEMINI.md} & 179 & 0.8\% & 83.2\% \\
\file{.claude/CLAUDE.md} & 150 & 0.7\% & 56.7\% \\
/\file{.cursorrules} & 94 & 0.4\% & 66.0\% \\
case variant (e.g. \file{agents.md}) at root & 47 & 0.2\% & 8.5\% \\
/\file{.windsurfrules} & 20 & 0.1\% & 95.0\% \\
\textbf{any of the above} & \textbf{6,652} & \textbf{29.1\%} &  \\
\bottomrule
\end{tabular}

\caption{Agent instruction formats in the \nPopulation{} repositories, listed one level deep. The last column
is the share of repositories with the format that also have a root \file{AGENTS.md}.}
\label{tab:formats}
\end{table}

\FloatBarrier
\subsection{The two file names}\label{sec:app-filenames}

\paragraph{Relations.} \Cref{tab:relations} gives the relation of the two files in every repository with an
agent file. When one file defers to the other, \file{CLAUDE.md} defers to \file{AGENTS.md} in \nDefers{}
repositories and the reverse holds in \nReverse.
\begin{itemize}
  \item Symbolic links point from \file{CLAUDE.md} to \file{AGENTS.md} in \nSymCA{} repositories, against
        \nSymAC{} the other way.
  \item \nImports{} use an \file{@AGENTS.md} import.
  \item The ratio of deference stays near eight to one at every thin-import threshold from \thresholdLow{} to
        \thresholdHigh{} characters. The number of repositories with two files of their own content ranges
        from \nTwoRealAtHighThreshold{} to \nTwoRealAtLowThreshold{} over the same thresholds.
\end{itemize}

\begin{table}[htbp]
\centering\small
\begin{tabular}{>{\raggedright\arraybackslash}p{0.42\linewidth}rr}
\toprule
Relationship & Repos & Share \\
\midrule
\file{AGENTS.md} only & 2,549 & 42.4\% \\
\file{AGENTS.md} only (symlink to another file) & 17 & 0.3\% \\
\file{CLAUDE.md} only & 1,040 & 17.3\% \\
\file{CLAUDE.md} only (symlink to another file) & 12 & 0.2\% \\
both: \file{CLAUDE.md} is a symlink to \file{AGENTS.md} & 629 & 10.5\% \\
both: \file{AGENTS.md} is a symlink to \file{CLAUDE.md} & 155 & 2.6\% \\
both: symlinks to the same other file & 28 & 0.5\% \\
both: \file{CLAUDE.md} is a thin \file{@AGENTS.md} import ($\leq$ 300 chars) & 771 & 12.8\% \\
both: \file{CLAUDE.md} imports \file{@AGENTS.md} and adds its own content & 101 & 1.7\% \\
both: \file{CLAUDE.md} is a short pointer to \file{AGENTS.md} & 140 & 2.3\% \\
both: \file{AGENTS.md} only imports or points to \file{CLAUDE.md} & 27 & 0.4\% \\
both: identical copies & 40 & 0.7\% \\
both: separate files, \file{CLAUDE.md} mentions \file{AGENTS.md} & 171 & 2.8\% \\
both: separate files, \file{AGENTS.md} mentions \file{CLAUDE.md} & 64 & 1.1\% \\
both: separate files & 262 & 4.4\% \\
\bottomrule
\end{tabular}

\caption{How the two files relate in the \nAdopters{} repositories with an agent file. The rules are applied
in order (\cref{sec:population}).}
\label{tab:relations}
\end{table}

\paragraph{Content moving between file names.} In \nMigrations{} repositories that started with \file{CLAUDE.md},
that file was later turned into a bridge to \file{AGENTS.md}; in \nMigOwn{} of them it had content of its own
before. \pctMigSameCommit{} of these conversions happened in the commit that created \file{AGENTS.md}, so the
content typically moved to \file{AGENTS.md} as the bridge was made. Conversions are less often AI-marked
(\pctMigAI) than the other \file{CLAUDE.md} commits in the same repositories (\pctMigAIBase).

\paragraph{Order and shift over time.} In the \nBoth{} repositories with both files, the files arrived in
the same commit in \pctSameCommit, \file{CLAUDE.md} came first in \pctClaudeFirst, and \file{AGENTS.md}
came first in \pctAgentsFirst. \Cref{tab:firstfile} gives the first file of new adopters by period.
\begin{itemize}
  \item \textbf{Renames.} Path history does not follow renames. In a random sample of \nRenameSample{}
        repositories whose first file is \file{AGENTS.md}, \pctRenamedFromClaude{} (95\% CI \ciRenamed) had
        in fact started with \file{CLAUDE.md}. Even at the upper bound for the latest period, the
        \file{CLAUDE.md}-first share there would rise to at most \pctClaudeFirstLateMax.
  \item \textbf{Claude Code's support for \file{AGENTS.md}.} Claude Code v2.1.277, released on
        \fallbackDate, began reading \file{AGENTS.md} when no \file{CLAUDE.md} is present
        \citep{anthropic2026changelog,anthropic2026memory}. In the \fallbackWindowDays{} days before it,
        \pctFallbackBefore{} of new adopters started with \file{AGENTS.md}; in the \fallbackWindowDays{} days
        after, \pctFallbackAfter. This fits the existing trend and does not show an effect of the release.
\end{itemize}

\begin{table}[htbp]
\centering\small
\begin{tabular}{lrrrr}
\toprule
Period & New adopters & First file \file{AGENTS.md} & First file \file{CLAUDE.md} & Both at once \\
\midrule
2025-01 -- 2025-08 & 536 & 23.9\% & 73.7\% & 2.4\% \\
2025-09 -- 2025-12 & 759 & 54.2\% & 36.8\% & 9.1\% \\
2026-01 -- 2026-04 & 2,178 & 48.3\% & 35.9\% & 15.8\% \\
2026-05 -- 2026-10-16 & 2,532 & 60.0\% & 17.1\% & 22.9\% \\
\bottomrule
\end{tabular}

\caption{New adopters by the date of their first surviving agent file, and which file came first.}
\label{tab:firstfile}
\end{table}

\FloatBarrier
\subsection{Origins per instruction set}\label{sec:app-origins}

\Cref{tab:origins} lists the traces behind \cref{sec:rq4}. Text traces cover all \nDocuments{} instruction sets;
history traces cover the \nOriginHist{} whose text-carrying files have a history of their own. A generator
signature is the opening line of \texttt{claude /init} (verbatim or with the agent's name changed) or the heading
set of Codex's \texttt{/init}. A tool-inserted block is recognised by the delimiter that the tool writes. A shared
paragraph has at least \paraMinChars{} characters and appears verbatim in the files of at least \sharedOwners{} owners; generator lines
and licence text are not counted. Near-duplicates have a MinHash Jaccard similarity of at least \nearDupJaccard.

\begin{table}[htbp]
\centering\small
\caption{Traces of origin per instruction set. Shares of all instruction sets for text traces and of those with
a file history for history traces. Traces identify signatures, shared text or marked commits; they do not establish authorship.}
\label{tab:origins}
\begin{tabular}{>{\raggedright\arraybackslash}p{0.42\linewidth}rr}
\toprule
Trace & Repos & Share \\
\midrule
generator signature (\file{claude /init} line or variant, Codex \file{/init} headings) & 948 of 5,998 & 15.8\% \\
of which \file{claude /init}'s opening line, verbatim & 692 & 11.5\% \\
tool-inserted block (Next.js, Turborepo, Nx, Laravel Boost, beads) & 93 & 1.6\% \\
paragraph shared verbatim with files of $\geq$ 2 other owners & 205 & 3.4\% \\
of which with no generator signature or tool block & 104 & 1.7\% \\
whole file near-duplicate of another owner's file & 60 & 1.0\% \\
\textbf{any text trace} & \textbf{1,166 of 5,998} & \textbf{19.4\%} \\
text file created in an AI-marked commit & 1,054 of 5,942 & 17.7\% \\
AI-marked commit after creation & 2,163 of 5,942 & 36.4\% \\
any AI-marked commit & 2,475 of 5,942 & 41.7\% \\
changed after creation (more than one commit) & 4,860 of 5,942 & 81.8\% \\
\file{CLAUDE.md} converted into a bridge to \file{AGENTS.md} & 396 of 5,998 & 6.6\% \\
\textbf{any text trace or AI-created} & \textbf{1,967 of 5,942} & \textbf{33.1\%} \\
\bottomrule
\end{tabular}

\end{table}

\FloatBarrier
\subsection{The prior codebook: our population against the reference study}\label{sec:app-stage-b}

\begin{table}[htbp]
\centering\footnotesize
\begin{tabularx}{\linewidth}{@{}>{\raggedright\arraybackslash}Xrrr>{\raggedright\arraybackslash}X@{}}
\toprule
Type & Ours, raw & \textbf{Ours, corrected} & \textbf{Reference} & Verdict \\
\midrule
AI Integration & 72.5\% & \textbf{71\% (59--92)} & \textbf{22\% (17--28)} & more common in ours \\
Documentation and References & 83.7\% & \textbf{72\% (62--85)} & \textbf{29\% (24--36)} & more common in ours \\
Development Process & 78.0\% & \textbf{91\% (82--100)} & \textbf{68\% (62--74)} & more common in ours \\
DevOps & 38.9\% & \textbf{29\% (23--36)} & \textbf{17\% (12--22)} & more common in ours \\
Security & 41.8\% & \textbf{28\% (22--33)} & \textbf{17\% (13--23)} & more common in ours \\
UI/UX & 24.3\% & \textbf{20\% (14--32)} & \textbf{6\% (4--10)} & more common in ours \\
Project Management & 17.4\% & \textbf{20\% (13--44)} & \textbf{5\% (3--9)} & more common in ours \\
Testing & 80.4\% & \textbf{67\% (51--75)} & \textbf{83\% (77--87)} & less common in ours \\
Build and Run & 78.5\% & \textbf{49\% (27--61)} & \textbf{71\% (64--76)} & less common in ours \\
System Overview & 65.1\% & 48\% (37--56) & 57\% (50--63) & similar (intervals overlap) \\
Architecture & 73.5\% & 70\% (63--75) & 71\% (64--76) & similar \\
Implementation Details & 74.3\% & 69\% (54--80) & 71\% (64--76) & similar \\
Configuration \& Environment & 63.3\% & 39\% (25--50) & 43\% (37--50) & similar \\
Maintenance & 53.9\% & 84\% (55--100) & 51\% (44--57) & unclear (very wide interval) \\
Debugging & 31.0\% & 17\% (6--27) & 23\% (18--29) & similar \\
Performance & 32.3\% & 15\% (8--21) & 13\% (9--18) & similar \\
\bottomrule
\end{tabularx}

\caption{Share of files with each instruction type of \citet{chatlatanagulchai2026agentreadmes}: ours raw, ours corrected for over-labelling with the Rogan--Gladen estimator (95\% intervals), and the reference study's human coding of its \file{AGENTS.md} and \file{CLAUDE.md} files. The samples differ in frame and date, so a difference shows that the populations differ, not why.}
\label{tab:stage-b}
\end{table}

\FloatBarrier
\clearpage
\subsection{Project types}\label{sec:app-types}

\begin{table}[htbp]
\centering\small
\caption{Project types of all adopters and all classified controls (model-classified by Claude Haiku 5.5 from
repository metadata; indicative, see \cref{sec:rq5}).}
\label{tab:e-types}
\begin{tabular}{lrr}
\toprule
Derived type & Adopters \% (n = 6,006) & Controls \% (n = 5,920) \\
\midrule
data / ML / AI (any AI focus) & 32.2 & 18.6 \\
library / framework & 23.0 & 32.0 \\
application & 22.3 & 22.3 \\
CLI / developer tool & 10.7 & 13.2 \\
infrastructure / DevOps & 8.0 & 6.0 \\
docs / learning & 2.6 & 6.0 \\
other or unclear & 1.2 & 1.7 \\
\bottomrule
\end{tabular}

\end{table}

\FloatBarrier
\subsection{Content and project characteristics}\label{sec:app-stage-e}

\begin{table}[htbp]
\centering\small
\caption{Kinds of statement per instruction set: Sonnet's codes on \nReposEone{} sample repositories, reweighted,
with 95\% bootstrap intervals; the last column is the corrected Haiku census.}
\label{tab:e-kinds}
\begin{tabular}{lrrr}
\toprule
Kind & \% of repos with $\geq$ 1 & Mean share of statements & Census (Haiku, corr.) \\
\midrule
rule & 98.2 (97.3--98.9) & 52.6 (50.9--54.4) & 53.0 \\
reference & 89.7 (87.8--91.6) & 17.6 (16.5--18.8) & 16.2 \\
description & 88.6 (86.6--90.5) & 29.8 (28.3--31.3) & 30.7 \\
\bottomrule
\end{tabular}

\end{table}

\begin{table}[htbp]
\centering\small
\caption{Subject areas: share of repositories with at least one rule (95\% CI) and with any statement (the data of
\cref{fig:content}).}
\label{tab:e-areas}
\begin{tabular}{lrr}
\toprule
Area & \% with a rule (95\% CI) & \% with any statement \\
\midrule
code conventions and quality & 74.4 (71.4--77.3) & 77.0 \\
build, environment and delivery & 72.6 (69.4--75.7) & 87.6 \\
agent workflow and conduct & 67.3 (64.0--70.3) & 68.5 \\
documentation & 61.7 (58.4--65.1) & 77.3 \\
testing & 60.2 (56.8--63.4) & 76.7 \\
version control and collaboration & 58.0 (54.3--61.7) & 60.9 \\
project knowledge & 54.3 (50.8--57.6) & 86.2 \\
technical domains & 53.2 (49.6--56.8) & 61.6 \\
agent tooling & 22.6 (19.3--26.0) & 29.1 \\
\bottomrule
\end{tabular}

\end{table}

\begin{table}[htbp]
\centering\small
\caption{Exploratory models: tests per predictor that pass Benjamini--Hochberg within the exploratory family (\nTestsE{}
tests). Content models use \nReposE{} instruction sets; the adopter-versus-control row uses \nAdoptersCompare{}
adopters and \nControlsCompare{} controls.}
\label{tab:e-predictors}
\begin{tabular}{lr}
\toprule
Predictor & Significant / tests \\
\midrule
commits in 12 months & 11 / 13 \\
age & 9 / 13 \\
project type (joint) & 6 / 13 \\
language (joint) & 6 / 13 \\
human authors & 4 / 13 \\
AI-marked share & 4 / 13 \\
stars & 3 / 13 \\
adopter vs control, by type & 6 / 8 \\
\bottomrule
\end{tabular}

\end{table}

\begin{table}[htbp]
\centering\small
\caption{Adopters and matched controls by project type (within star tier and language strata; controls reweighted
to the adopters' strata). Types are model-classified and indicative.}
\label{tab:e-control}
\begin{tabular}{lrrrr}
\toprule
Type & Adopters \% & Controls \% & Odds ratio (95\% CI) & p (FDR) \\
\midrule
data / ML / AI & 32.2 & 18.6 & 2.38 (2.15--2.64) & < 0.001 \\
LLMs / generative AI / agents (AI focus) & 27.9 & 13.3 & 2.76 (2.48--3.08) & < 0.001 \\
infrastructure & 8.6 & 6.6 & 1.38 (1.17--1.61) & < 0.001 \\
application & 21.9 & 21.9 & 1.00 (0.91--1.11) & 0.96 \\
developer tool & 10.9 & 13.6 & 0.77 (0.68--0.87) & < 0.001 \\
library / framework & 23.0 & 33.3 & 0.58 (0.53--0.64) & < 0.001 \\
docs / learning & 2.3 & 4.5 & 0.50 (0.39--0.62) & < 0.001 \\
\bottomrule
\end{tabular}

\end{table}

\begin{table}[htbp]
\centering\small
\caption{The human reviewer's blind check of Haiku's project types on \nSpotScored{} repositories.}
\label{tab:e-spot}
\begin{tabular}{lrrr}
\toprule
Field & PI agrees with Haiku & Wilson 95\% CI & Cohen's $\kappa$ (95\% CI) \\
\midrule
AI focus & 27 / 29 (93.1\%) & 78--98\% & 0.80 (0.44--1.00) \\
type & 16 / 29 (55.2\%) & 38--72\% & 0.43 (0.23--0.63) \\
derived seven-way type & 17 / 29 (58.6\%) & 41--74\% & 0.50 (0.30--0.70) \\
\bottomrule
\end{tabular}

\end{table}

\begin{table}[htbp]
\centering\small
\caption{Repositories with agent instructions by star tier (the data of \cref{fig:adoption}A).}
\label{tab:tiers}
\begin{tabular}{lrrrr}
\toprule
Stars & Repos & \file{AGENTS.md} & \file{CLAUDE.md} & Either \\
\midrule
100k+ & 89 & 52.8\% & 31.5\% & 59.6\% \\
10k--100k & 3,059 & 32.3\% & 21.5\% & 37.0\% \\
5k--10k & 3,087 & 26.1\% & 17.9\% & 31.2\% \\
1k--5k & 16,619 & 18.7\% & 13.2\% & 23.2\% \\
\bottomrule
\end{tabular}

\end{table}

\FloatBarrier
\subsection{Element taxonomies}\label{sec:app-elements}

All element categories for the four collaboration subjects, with Haiku's codes on the full corpus (uncorrected;
stance judged per element). Definitions and examples are released with the data.

\begin{table}[htbp]
\centering\footnotesize
\caption{Pull-request descriptions: share of the \nPRFiles{} files with a rule on the subject that require or prohibit each element.}
\label{tab:full-pr}
\begin{tabular}{>{\raggedright\arraybackslash}p{0.42\linewidth}rr}
\toprule
Category & \% requiring & \% prohibiting \\
\midrule
Test plan / verification evidence & 27.2 & 4.4 \\
Use and complete the PR template & 24.2 & 1.6 \\
Linked issue and related PR references & 23.3 & 1.3 \\
Title format / type, tag or scope prefix & 19.3 & 1.7 \\
Summary of the change and its behavior effect & 17.8 & 0.1 \\
Screenshots, recordings and visual evidence & 15.6 & 1.1 \\
Motivation, problem statement and root cause & 12.9 & 0.2 \\
Concision, length limits and excluded noise & 9.3 & 4.8 \\
Breaking changes, compatibility and migration impact & 8.6 & 0.4 \\
Structure, formatting and language & 8.6 & 3.0 \\
PR creation workflow and agent authorship permission & 8.5 & 3.5 \\
Other & 7.7 & 1.4 \\
Design decisions, trade-offs and reviewer guidance & 7.4 & 0.2 \\
Title wording, length and style & 6.5 & 1.1 \\
General description quality & 6.2 & 0.4 \\
Labels, branch target and PR metadata & 5.8 & 0.4 \\
AI-use disclosure and agent attribution & 5.7 & 1.9 \\
Risks, limitations, unverified items and follow-ups & 5.4 & 0.3 \\
PR scope and size & 5.1 & 1.0 \\
Accuracy and kept in sync with the final diff & 4.7 & 1.0 \\
Release notes, changelog and changeset entries & 3.5 & 0.3 \\
\bottomrule
\end{tabular}

\end{table}

\begin{table}[htbp]
\centering\footnotesize
\caption{Commit messages: share of the \nCommitFiles{} files with a rule on the subject that require or prohibit each element.}
\label{tab:full-commit}
\begin{tabular}{>{\raggedright\arraybackslash}p{0.42\linewidth}rr}
\toprule
Category & \% requiring & \% prohibiting \\
\midrule
Conventional Commits / structured header format & 36.7 & 1.8 \\
Allowed commit types and when to use each & 23.5 & 0.6 \\
Subject mood, case and punctuation & 15.1 & 1.3 \\
Scope / component / area prefix & 14.9 & 0.8 \\
Body presence, layout and wrapping & 12.6 & 1.2 \\
Rationale and change description in the message & 12.6 & 2.0 \\
Issue, ticket or PR reference & 11.4 & 1.7 \\
Subject / header length limit & 10.0 & 0.0 \\
Follow project conventions, tooling and PR-title alignment & 6.6 & 0.2 \\
General clarity and quality & 6.5 & 0.5 \\
One logical change per commit & 6.3 & 0.1 \\
Sign-off, signing and human authorship & 5.9 & 1.9 \\
Message language & 5.0 & 0.1 \\
Other & 3.6 & 1.2 \\
AI / agent attribution & 3.5 & 7.4 \\
Breaking-change marker & 3.2 & 0.3 \\
Other trailers and control markers & 3.2 & 1.6 \\
Inline markup and mentions & 1.6 & 0.5 \\
Emoji / gitmoji prefixes & 1.3 & 0.6 \\
Process noise and non-public content & 0.3 & 2.1 \\
\bottomrule
\end{tabular}

\end{table}

\begin{table}[htbp]
\centering\footnotesize
\caption{Test writing: share of the \nTestFiles{} files with a rule on the subject that require or prohibit each element.}
\label{tab:test-elements}
\begin{tabular}{>{\raggedright\arraybackslash}p{0.42\linewidth}rr}
\toprule
Category & \% requiring & \% prohibiting \\
\midrule
Tests must accompany changes (and exemptions) & 28.2 & 1.9 \\
Frameworks, libraries, helpers and selectors & 19.5 & 2.2 \\
Test structure, granularity and conventions & 17.4 & 2.4 \\
Required scenarios, edge cases and error paths & 16.8 & 0.1 \\
Choice of test type / level & 15.5 & 1.7 \\
Regression test for bug fixes & 14.3 & 0.3 \\
Fixtures, factories and test data & 13.1 & 2.5 \\
Mocking and test doubles policy & 12.5 & 5.4 \\
Assertion strength and precision & 12.4 & 2.7 \\
Other & 12.1 & 2.8 \\
Determinism, timing and portability & 10.7 & 4.3 \\
Test file location and file naming & 10.4 & 0.7 \\
Test naming, descriptions and comments & 10.3 & 1.7 \\
Test isolation, shared state and cleanup & 9.1 & 2.0 \\
Test observable behavior, not implementation & 8.7 & 4.6 \\
Test-first / TDD workflow & 6.5 & 0.3 \\
Snapshots and golden files & 4.7 & 1.5 \\
General test quality & 4.6 & 0.3 \\
Coverage metrics and thresholds & 4.3 & 0.5 \\
Test worthiness / no trivial or tautological tests & 3.6 & 8.1 \\
No skipping, weakening or deleting tests & 3.1 & 4.9 \\
\bottomrule
\end{tabular}

\end{table}

\begin{table}[htbp]
\centering\footnotesize
\caption{Code review: share of the \nReviewFiles{} files with a rule on the subject that require or prohibit each element.}
\label{tab:full-review}
\begin{tabular}{>{\raggedright\arraybackslash}p{0.42\linewidth}rr}
\toprule
Category & \% requiring & \% prohibiting \\
\midrule
Review process, reviewers, tools and gating & 22.2 & 2.7 \\
Correctness, edge cases and error handling & 20.2 & 2.0 \\
Context gathering and review method & 17.7 & 1.5 \\
Project guidelines, checklists and repo-specific checks & 17.5 & 1.2 \\
Evidence, verification and confidence of findings & 14.5 & 1.5 \\
Review report format, verdict, tone and language & 14.5 & 1.0 \\
Test coverage and test quality & 13.0 & 1.5 \\
Handling received review feedback and threads & 13.0 & 2.0 \\
Architecture, simplicity and maintainability & 12.5 & 2.2 \\
Security and privacy & 12.2 & 2.0 \\
Finding content: location, impact and suggested fix & 11.2 & 0.7 \\
Severity levels and blocking vs optional & 11.0 & 1.7 \\
Scope: diff-introduced issues, scope creep and PR splitting & 10.2 & 4.2 \\
Backward compatibility and breaking changes & 9.5 & 0.7 \\
Style, naming, formatting and readability & 7.0 & 7.7 \\
Performance and efficiency & 6.0 & 0.5 \\
General review quality / follow the review process & 5.7 & 0.5 \\
Concurrency, resource leaks and lifecycle & 5.5 & 0.7 \\
Documentation, changelog and release notes & 5.2 & 0.5 \\
Agent permissions, posting and AI disclosure & 4.5 & 3.2 \\
Other & 1.7 & 1.2 \\
\bottomrule
\end{tabular}

\end{table}

\FloatBarrier
\subsection{Deviations from the analysis plan}\label{sec:app-deviations}

\begin{enumerate}
  \item The planned mixed logistic model for the paired hypothesis was replaced by a GEE logistic model with
        repository clusters; the mixed model was unstable at this size.
  \item The exploratory set differs from the prespecified list. File length uses stars only. The
        composition analysis is a main effect rather than an interaction. The per-language subject profiles
        were not run.
  \item The paired test first included repositories whose two files are identical copies, and repositories
        excluded by the activity criterion. It was corrected to the prespecified population before
        reporting; no other test was affected.
  \item Post hoc checks, labelled as such, were added after the results were seen:
        \begin{itemize}
          \item the AI-share hypothesis without the \nAIBanRepos{} repositories whose files prohibit AI
                attribution;
          \item single-predictor models for the collaboration hypothesis;
          \item the paired effect in percentage points, with a bootstrap interval.
        \end{itemize}
\end{enumerate}

\end{document}